\documentclass[aps,prx,twocolumn,showpacs,superscriptaddress,10pt]{revtex4-2}  
\usepackage{bbm}
\usepackage{epsfig}
\usepackage{epstopdf}
\usepackage{graphicx}
\usepackage{amsmath,amssymb}
\usepackage{amsmath,bm}
\usepackage{babel}

\usepackage{txfonts}
\usepackage{physics}
\usepackage{comment}
\usepackage{xcolor}
\usepackage{lipsum}
\usepackage[caption=false]{subfig}
\usepackage{lineno,blindtext}
\usepackage{diagbox, eqparbox, hhline}
\usepackage{soul}
\usepackage{xcolor}
\usepackage[colorlinks,citecolor=blue,linkcolor=blue,urlcolor=blue,]{hyperref}

\begin{document}
	
	\title{{Kinetically constrained cavity QED:\\ from blockaded ferromagnetism to long-range quantum scars}}

	\author{Hossein Hosseinabadi}
	\email{hossein@pks.mpg.de}
	\affiliation{Institute of Physics, Johannes Gutenberg University Mainz, 55099 Mainz, Germany}

    \author{Riccardo J. Valencia-Tortora}
     \affiliation{Institute of Physics, Johannes Gutenberg University Mainz, 55099 Mainz, Germany}

    \author{Aleksandr N. Mikheev}
    \affiliation{Institute of Physics, Johannes Gutenberg University Mainz, 55099 Mainz, Germany}
    \affiliation{Department of Physics, University of Konstanz, Universit{\"a}tsstra{\ss}e 10, 78464 Konstanz, Germany}

    \author{Darrick E. Chang}
    \affiliation{ICFO—Institut de Ci{\`e}ncies Fot{\`o}niques, The Barcelona Institute of\\ Science and Technology, 08860 Castelldefels, Spain}
    \affiliation{ICREA—Instituci{\'o} Catalana de Recerca i Estudis Avan{\c c}ats, 08015 Barcelona, Spain}

    \author{Johannes Zeiher}
     \affiliation{Max-Planck-Institut f\"{u}r Quantenoptik, 85748 Garching, Germany}
     \affiliation{Munich Center for Quantum Science and Technology (MCQST), 80799 Munich, Germany}
     \affiliation{Fakultät für Physik, Ludwig-Maximilians-Universit\"{a}t, 80799 Munich, Germany}

    \author{Roderich Moessner}
	\affiliation{Max Planck Institute for the Physics of Complex Systems, N\"othnitzer Str.~38, 01187 Dresden, Germany}
	
	\author{Jamir Marino}
	\affiliation{Institute of Physics, Johannes Gutenberg University Mainz, 55099 Mainz, Germany}
    \affiliation{Department of Physics, The State University of
New York at Buffalo, NY 14260, USA}
	
	\begin{abstract}
		Rydberg–cavity systems are emerging as promising platforms for quantum simulation and quantum information processing. These hybrid architectures combine two complementary interaction mechanisms: cavity photons mediate collective long-range couplings, while Rydberg excitations generate strong short-range interactions. Together, they offer a   setting for engineering many-body phases characterized by a hierarchy of interactions across widely different length scales. In this work, we introduce a minimal and scalable model for such systems. Focusing on the strong Rydberg blockade regime, we restrict the Hilbert space to the subspace enforced by the blockade, yielding a kinetically constrained long-range model in one spatial dimension. This approach both captures the physics of Rydberg–cavity experiments in the regime of strong Rydberg interactions and provides a conceptually transparent framework for studying the interplay of long-range and short-range interactions. At equilibrium, in addition to paramagnetic and N\'eel-ordered phases, the system supports a blockaded ferromagnetic/superradiant phase, distinct from the conventional superradiant phase. Out of equilibrium, we identify long-range quantum many-body scars, which are atypical nonthermal eigenstates that evade the eigenstate thermalization hypothesis, and giving rise to slow entanglement growth. In contrast to the linear-in-time entanglement growth characteristic of short-range scarred models, these long-range scars exhibit logarithmic entanglement dynamics. Our results establish a minimal yet versatile framework for Rydberg–cavity systems, and provide a stepping stone for future theoretical and experimental studies of this frontier platform in quantum many-body physics.
	\end{abstract}
	
	\maketitle

	\section{Introduction}
	\label{sec:intro}

		Cold-atom systems are established as versatile quantum simulators, offering unprecedented control over interactions and geometry~\cite{Bloch2008,gross2017quantum,PRXQuantum.2.017003}. Among their many incarnations, two directions stand out for their ability to engineer strongly interacting quantum matter: cavity quantum electrodynamics (QED), where long-range photon-mediated interactions dominate, and Rydberg atom arrays, which realize programmable spin models with strong short-range couplings. Together, these complementary platforms have enabled the exploration of collective phases of matter, novel dynamical regimes, and new paradigms of light–matter interaction.
		
			\begin{figure}[!t]
			\centering
			\includegraphics[width=0.98\linewidth]{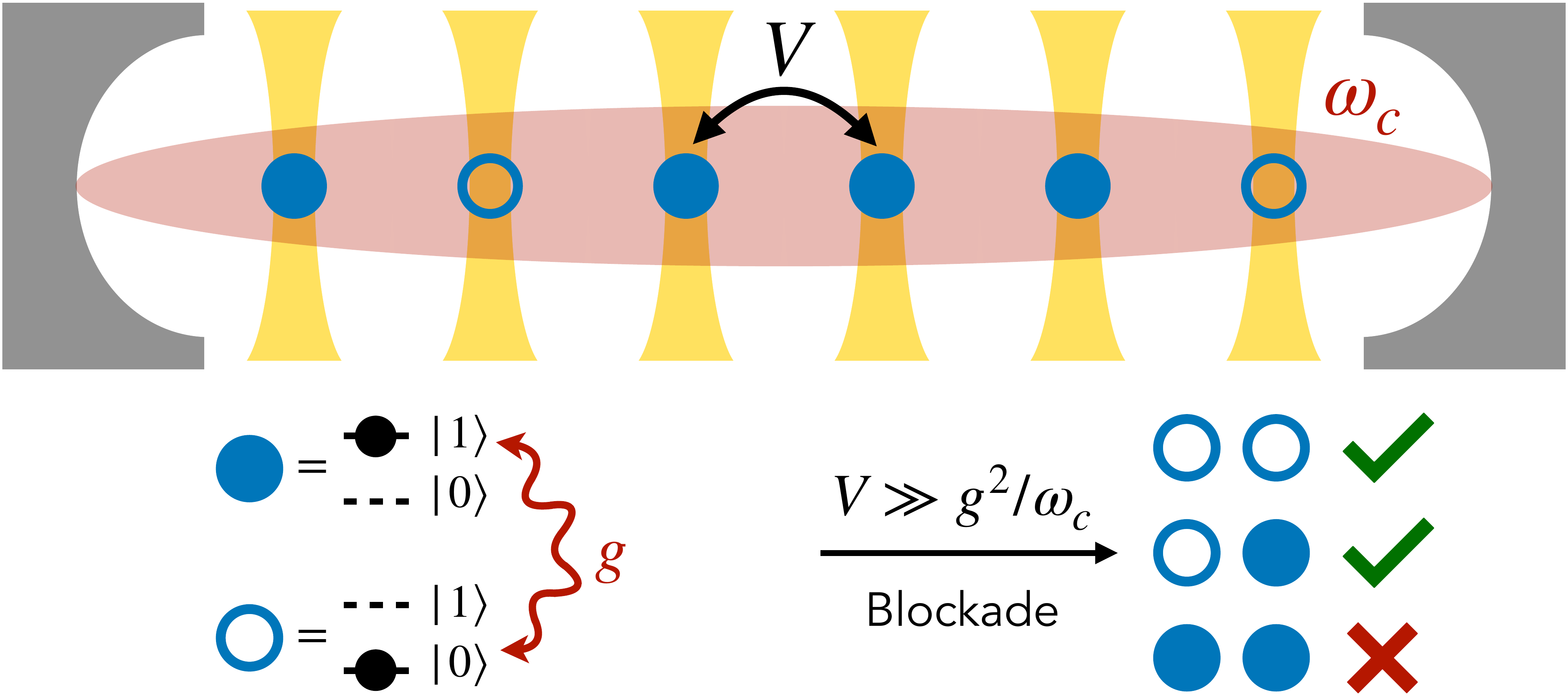}
			\caption{\textbf{Schematics of the model.} A chain of two-level atoms is placed inside an optical cavity and suspended by optical tweezers (yellow shades). The cavity mode (red shade) mediates long-range interactions that generate hybridization between atomic ground and excited states across the chain (red wavy line). In addition, atoms experience short-range Rydberg interactions that energetically penalize configurations with simultaneously excited nearest neighbors (blockade). The interplay of short-range and long-range interactions leads to distinct physics, both in and out-out equilibrium.}
			\label{fig:cartoon}
		\end{figure}

		Cavity QED provides a natural setting for simulating long-range interacting systems. When atomic ensembles are placed between high-finesse mirrors, photons mediate effective couplings that extend across the entire system and can be tuned via external driving~\cite{Mivehvar_Piazza_Donner_Ritsch_2021}. A hallmark example is the Dicke superradiant phase transition~\cite{Dicke1954,HEPP1973360,Kirton_Dicke2019}, where light and matter become coherently locked into a collective state. Experiments have realized Dicke superradiant using Bose–Einstein condensates in cavities~\cite{Baumann2010,Baumann2011,Landig2016,Leonard2017}, with recent extensions to fermionic gases~\cite{Piazza_Strack_2014,Chen_Yu_Zhai_2014,Keeling_Bhaseen_Simons_2014,Helson2023,Zwettler2025}, associative memory models~\cite{Marsh2021}, and spin-glass dynamics~\cite{Gopalakrishnan2011,strack_dickeglass2011,Marsh2024,Kroeze2025,marsh2025multimode,Hosseinabadi2024short,Hosseinabadi2024long}.
		
		Rydberg atom arrays, by contrast, realize short-range interacting spin models through the strong and tunable interactions of atoms in highly excited Rydberg states~\cite{browaeys2020many}. This architecture allows for flexible geometries and interaction graphs, enabling the study of ordered magnetic phases~\cite{Lesanovsky_Rydberg2011,Ebadi2021,Scholl2021,Chen2023} as well as exotic states such as topological matter and quantum spin liquids~\cite{Leseleuc2019,Semeghini2021}. Beyond equilibrium phases, Rydberg systems also exhibit striking non-equilibrium dynamics. Most notably, they host quantum many-body scars~\cite{Bernien_Rydberg2017,Bluvstein2021}, atypical nonthermal eigenstates with low entanglement that evade the eigenstate thermalization hypothesis and give rise to long-lived coherent oscillations~\cite{Turner_scars2018,Ho_scars2019,Serbyn_scars2021,Moudgalya_2022, valencia2024rydberg,
valencia2022kinetically,PhysRevLett.129.020601}.

		The complementary strengths of cavity QED and Rydberg arrays motivate hybrid platforms that combine the two, opening new opportunities for quantum simulation. One realization of such systems involves ensembles of Rydberg atoms, known as Rydberg superatoms, embedded in optical cavities, which have been studied primarily in the context of quantum information processing~\cite{Clark_CavityRydberg2020,Stolz_cavityRydberg2022,Vaneecloo_cavityRydberg2022}. A more recent and rapidly growing direction is the integration of ordered Rydberg arrays into cavities, which promises to merge the advantages of global photon-mediated interactions with the programmable short-range couplings of Rydberg systems. Although experimental progress on atomic arrays in cavities is still at an early stage~\cite{Deist_cavityArray1_2022,Deist_cavityArray2_2022,Yan_cavityArray2023,Hu_TweezersCavity2025}, a growing body of theoretical work has predicted a wealth of novel phenomena. These include phases with intertwined superradiant and antiferromagnetic or ferromagnetic order~\cite{Gelhausen_Buchhold_Rosch_Strack_2016,Puel_Macri_2024}, the emergence of meson–polariton excitations~\cite{Bacciconi_Xavier_Marinelli_Bhakuni_Dalmonte_2025}, and long-lived prethermal regimes of coupled light and matter~\cite{mikheev2025prethermalization}. These proposals highlight the potential of hybrid cavity-Rydberg systems to access correlated regimes of light and matter that lie beyond the reach of either platform alone.

		Future experimental progress in hybrid cavity–Rydberg systems will need to be closely guided by theory. To date, theoretical studies have largely relied either on exact treatments restricted to small system sizes~\cite{Puel_Macri_2024,Bacciconi_Xavier_Marinelli_Bhakuni_Dalmonte_2025} or on approximate approaches such as mean-field and semiclassical methods~\cite{mikheev2025prethermalization,hosseinabadi2025TWA}, whose validity must be carefully assessed. This situation underscores the necessity of advancing along two complementary directions. First, the development of analytical and numerical techniques capable of accurately capturing the physics of cavity–Rydberg platforms. Second, the identification of minimal models that retain the essential features of these systems in certain regimes while remaining analytically tractable or amenable to numerically exact solutions to larger system sizes.

		In this work, we contribute to the second direction outlined above by introducing a minimal model of a Rydberg atom array coupled to an optical cavity (Fig.~\ref{fig:cartoon}). By focusing on the strong Rydberg blockade regime, we can exclude states with simultaneously excited neighboring atoms, and thereby work within a reduced Hilbert space, allowing to reach   larger system sizes.  The  latter would require treating both the atomic and photonic Hilbert spaces explicitly, which significantly restricts the accessible system sizes and makes finite-size effects particularly severe in presence of long-range interactions. Beyond its direct relevance for Rydberg–cavity experiments, the model is of intrinsic theoretical interest, as it provides one of the simplest, yet rich, frameworks to study the interplay of short-range and long-range interactions.
		
		Owing to the interplay between the strong Rydberg blockade and cavity-mediated interactions, our model realizes a kinetically constrained long-range system. As a result, its properties combine features characteristic of both kinetically constrained models, most notably the PXP model~\cite{Lesanovsky_Rydberg2011,Turner_scars2018,Ho_scars2019,Serbyn_scars2021}, and conventional long-range interacting systems~\cite{Ribeiro_LMG2007,Buyskikh_EntGrowthLongrange2016,Pappalardi_EntGrowthLongrange2018,Lerose_SlowGrowth2020,defenu_rmp2023,marino2022universality}. At the level of equilibrium phases, we find that the ground state can be tuned between a paramagnetic phase, a N\'eel-ordered phase, and most notably, a blockaded ferromagnetic/superradiant phase. Out of equilibrium, the system hosts  {long-range quantum many-body scars}, which manifest in the absence of thermalization and slow entanglement growth. Remarkably, the entanglement entropy in these scarred dynamics grows logarithmically in time~\cite{Pappalardi_EntGrowthLongrange2018}, in contrast to the linear growth observed in short-range scarred systems~\cite{Turner_scars2018,Ho_scars2019}.
        
        Our results provide a stepping stone for future studies of cavity–Rydberg systems, which represent one of the most promising frontiers for realizing novel quantum many-body phenomena.  {More broadly, our work establishes a route to realizing kinetically constrained dynamics in long-range interacting quantum systems, thereby opening a new direction for exploring many-body physics at the intersection of cavity QED and constrained dynamics. We further elaborate on the potential of this novel research area in the Concluding section.}

    \section{Summary of the Results}
    \label{sec:summary}
    In Section~\ref{sec:model}, we introduce our setup, consisting of a chain of atoms coupled to Rydberg states and the field of an optical cavity~(Fig.~\ref{fig:cartoon}), realizing a many-body system with competing short-range and long-range interactions. The short-range component arises from the Rydberg interactions, which energetically suppress simultaneous excitations of neighboring atoms, while the long-range interactions are mediated by the collective coupling of atoms to a common cavity mode. In the absence of Rydberg interactions, the physics of the system is very well captured by mean-field theory, as long-range interactions tend to suppress effects of quantum fluctuations~\cite{LIPKIN1965188,Kirton_Dicke2019,defenu_rmp2023}. The central question is how the strong short-range interactions modify the physics compared to  conventional cavity QED frameworks.

    By focusing on the regime of strong Rydberg interactions, in Section~\ref{sec:PXP2_derivation} we derive an effective description of the system constrained to the subspace allowed by the blockade. 
    We show that the resulting effective Hamiltonian, referred to as (PXP)$^2$, can be expressed in terms of the square of the well-known PXP Hamiltonian, augmented by an additional transverse field term. Alternatively, the model can be interpreted as a Dicke Hamiltonian projected onto the blockade subspace. Due to this duality in the structure of the model, its behavior, both at equilibrium and out of equilibrium, exhibits an interplay of the physics characteristic of the PXP and Dicke models.

    Using this effective theory, we first discuss the ground state phases of the system in Section~\ref{sec:equilibrium}. The model exhibits three distinct phases depending on the magnitude and sign of the transverse field. For large positive fields, the ground state is paramagnetic, with all atoms remaining in their ground states. In the opposite limit, for large negative fields, the system favors a N\'eel-ordered configuration with a period-2 density-wave profile. Between these two extremes, at intermediate values of the transverse field, the ground state becomes ferromagnetic or, with respect to the cavity mode, superradiant. This phase breaks the spin $Z_2$ symmetry while preserving translation symmetry, a surprising feature given the presence of (infinitely) strong Rydberg blockade. As a result, the ground state cannot be approximated by a classical product state, even in the limit of vanishing transverse field. This ordered state is strongly affected by quantum fluctuations induced by the blockade, which significantly 
'dress' the classical product state ansatz. A hallmark of this effect is a  magnetization that exceeds any classical value allowed by the blockade constraint.

    Next, we analyze the spectrum of low-energy excitations in Section~\ref{sec:excitations}. We numerically compute the dispersion of the lowest-lying modes and find that, in the paramagnetic phase, the spectrum is dispersive at finite momenta, while the zero-momentum mode remains energetically isolated by a finite gap below the continuum. This gap closes at the transition into the ferromagnetic phase, signaling the onset of spontaneous symmetry breaking. To understand these features, in Section~\ref{sec:field_theory} we develop a field-theoretic formalism that qualitatively captures the structure of the excitation spectrum. This analysis shows that the finite-momentum dispersive modes originate from the Rydberg blockade, while the isolated zero-momentum mode is a consequence of the long-range cavity-mediated interactions.

    We then proceed to investigate the non-equilibrium behavior of the system in Section~\ref{sec:dynamics}. Motivated by the connection to the PXP model, we examine the spectrum for signatures of quantum many-body scars (QMBS). Through spectral analysis, we show that, similar to the PXP case, the spectrum is predominantly chaotic but contains a set of atypical scarred eigenstates, violating volume-law entanglement expected for chaotic systems. These scarred states exhibit strong overlap with initial product states featuring short-period density-wave order, most notably the N\'eel state.

    In Section~\ref{sec:ent_dynamics}, to address whether these non-thermal eigenstates have an impact on dynamics, as they are just a polynomial number compared to the exponentially abundant number of thermal states, we analyze the entanglement dynamics following quantum quenches from various product states. Remarkably, we find that the half-cut entanglement entropy exhibits logarithmic growth in time across all initial states, regardless of their overlap with scarred eigenstates. However, states with stronger overlap with QMBS show slower, though still logarithmic, entanglement production. These states include density-wave configurations, which have a periodic distribution of excited atoms across the chain. We demonstrate that this slower growth can be understood within the framework of long-range interacting systems: by partitioning the chain into sublattices and neglecting the initially unoccupied sublattices, the dynamics of the remaining subsystem can be effectively described by a long-range Ising model without projectors. In this picture, the growth of entanglement is explained in terms of the production of collective excitations (magnons). For density-wave states, the reduced phase space for magnon creation due to projectors leads to suppressed entanglement growth compared to the vacuum state. This mechanism also explains why the period-three density-wave state exhibits slower entanglement growth than the period-two N\'eel state, an observation that contrasts with the behavior observed in the PXP model, where longer period density-wave states thermalize more efficiently.
    
    We also explore the impact of the transverse field on entanglement dynamics in Section~\ref{sec:ent_dyn_finite_D} and observe a striking asymmetry between positive and negative field values. For positive fields, entanglement growth is monotonically suppressed, consistent with the system becoming increasingly polarized. In contrast, for negative field values, entanglement production initially increases, reaching a pronounced maximum before decreasing again at larger negative fields. We show that this non-monotonic behavior arises from a resonance phenomenon: certain terms in the Hamiltonian effectively act as a drive on the system, and the resonance occurs when the drive frequency matches the excitation energy. This resonance condition coincides with the point of maximal entanglement growth, providing a clear dynamical interpretation of the observed peak.

    Finally, we briefly address the experimental feasibility of the model in Section~\ref{sec:experimental}, arguing that the omission of dissipation is well justified for experimentally relevant timescales, where coherent dynamics dominate.

    \section{Model}\label{sec:model}

    The goal of this paper is to put forward a minimal model of competing short-range and long-range interactions in the regime of strong interactions. This model can be realized by an array of Rydberg atoms held by optical tweezers inside of an optical cavity (Fig.~\ref{fig:cartoon}). The atoms are collectively coupled to a single cavity mode, leading to an effective long-range interaction between the atoms. While this Dicke-Ising model captures the essential features of Rydberg atoms embedded in optical cavities, it remains too complex to allow for exact solutions. To facilitate analytical and numerical progress, we focus on the regime of strong Rydberg interactions and project the model onto the subspace defined by the Rydberg blockade constraint. This projection yields an effective theory, whose properties will be investigated in detail throughout the remainder of this work. We focus on the one-dimensional case, which is most relevant to current experiments on Rydberg-dressed atoms in cavities.

    \subsection{The Dicke-Ising Hamiltonian}
    We consider a chain of two-level atoms of length $L$ inside an optical cavity (Fig.~\ref{fig:cartoon}). The atoms interact with each other by coupling to Rydberg states and with a single cavity mode, as described by the following Hamiltonian
    \begin{equation}\label{eq:H0}
        H = H_\mathrm{a} + H_\mathrm{c} + H_\mathrm{ac}.
    \end{equation}
    $H_a$ is the atomic part given by
    \begin{equation}
        H_\mathrm{a} = \Delta \sum_{i=1}^L \sigma^z_i + V\sum_{\langle i j \rangle} n_i n_j,
    \end{equation}
    where $\sigma^\alpha_i$ are the Pauli matrices satisfying $[\sigma^\alpha,\sigma^\beta]=i\epsilon_{\alpha \beta \gamma}\sigma^\gamma$, and $\Delta$ is a local field. $V>0$ is the strength of Rydberg interactions, which are approximated by the nearest-neighbor coupling of the excited state occupation, $2n_i\equiv 1+\sigma^z_i$. $H_\mathrm{c}$ describes a free bosonic mode
    \begin{equation}
        H_\mathrm{c} = \omega_c a^\dagger a,
    \end{equation}
    where $\omega_c$ is the cavity detuning. The atom-cavity interaction is given by
    \begin{equation}\label{eq:H_ac}
        H_\mathrm{ac}=-\frac{g}{\sqrt{L}}\sum_{i=1}^L (a+a^\dagger) \,\sigma^x_i,
    \end{equation}
    where $g$ is the coupling strength, which is rescaled with $L$ to keep the total energy extensive. In general, incoherent processes, such as atomic decay and cavity loss, are also present in experiments. We neglect these effects in our analysis, assuming that their effect is small during the runtime of experiments.

    The model in Eq.~\eqref{eq:H0} is well understood in two limiting cases. For $g = 0$, the spins and the cavity decouple, and the spin sector reduces to an Ising chain with antiferromagnetic nearest-neighbor interactions in a longitudinal field $h_\mathrm{ex} = \Delta + V/2$. In this limit, the energy spectrum can be obtained exactly, as all terms in the Hamiltonian commute. The opposite limit, $V = 0$, corresponds to the Dicke model, which has been extensively studied in atomic physics, both theoretically and experimentally~\cite{Kirton_Dicke2019,Mivehvar_Piazza_Donner_Ritsch_2021}. Here, the Hamiltonian has permutation symmetry under the exchange of spin indices and can be expressed in terms of collective spin operators $S^\alpha = \sum_i \sigma^\alpha_i$. This symmetry greatly reduces the numerical cost of exact diagonalization, enabling simulations of systems with hundreds of spins~\cite{Kirton_Dicke2019}. Furthermore, the mean-field (MF) solution of the Dicke model becomes increasingly accurate as we approach the thermodynamic limit, providing analytical insight into the system's behavior both in and out of equilibrium. We discuss the properties of both of these limiting cases in more detail in Section~\ref{sec:equilibrium}, in connection with the behavior of $H$ in more general parameter regimes.

    Away from the two limiting cases discussed above, we encounter the same challenges typical of many-body Hamiltonians: the Hilbert space grows exponentially with the system size, making it difficult to study large systems. Ref.~\cite{Gelhausen_Buchhold_Rosch_Strack_2016} studied a similar model using a MF treatment of both long-range and short-range interactions, which was later improved in Ref.~\cite{mikheev2025prethermalization} by systematically incorporating fluctuations using field-theoretic methods.  Recently, several works have investigated variants of $H$, focusing on regimes where the Ising interaction ($V$) is ferromagnetic~\cite{Puel_Macri_2024}, or where it is antiferromagnetic but combined with a staggered field. The latter case can be mapped to a ferromagnetic interaction in a uniform field via a $\pi$-rotation of the spins on one of the sublattices~\cite{Bacciconi_Xavier_Marinelli_Bhakuni_Dalmonte_2025}.

    In this work, we focus on the unexplored regime of strong Rydberg interactions, where $V$ is largest energy scale in Eq.~\eqref{eq:H0}. In this strong Rydberg blockade limit, configurations with simultaneously excited neighboring sites are energetically forbidden. As a result, the Hilbert space can be effectively projected onto the subspace allowed by the blockade, leading to a kinetically-constrained long-range system. A technical advantage of the projection is that the effective model can be exactly diagonalized for larger system sizes.

    \subsection{Effective Theory in the Strong Blockade Regime}\label{sec:PXP2_derivation}

    The strong blockade theory can be constructed following Refs.~\cite{Lesanovsky_Rydberg2011,Lesanovsky_Fibonacci2012}, by introducing projectors onto the low-energy subspace that excludes configurations with simultaneously excited adjacent sites, such that
    \begin{equation}\label{eq:blocakde}
        n_i n_{i+1} = 0.
    \end{equation}
    Among the terms in the Hamiltonian, only the atom-cavity interaction in Eq.~\eqref{eq:H_ac} requires this projection. In a chain geometry, a spin-flip operator $\sigma^x_i$ is allowed to act only if both neighboring sites are unoccupied. The corresponding projection is given by
\begin{equation}\label{eq:sx_project}
    \sigma^x_i \to \tilde{\sigma}^x_i = P_{i-1} \sigma^x_i P_{i+1},
\end{equation}
where $P_i = 1 - n_i$ projects onto the atomic ground state at site $i$. Applying this projection, the atom-cavity coupling becomes
\begin{equation}\label{eq:H_tilde_ac}
    H_\mathrm{ac} \to \tilde{H}_\mathrm{ac} = -\frac{g}{\sqrt{L}} \sum_i (a + a^\dagger)\, \tilde{\sigma}^x_i.
\end{equation}
In addition to the strong blockade regime, we work in the adiabatic limit, where the cavity mode evolves much faster than the atomic excitations within the projected subspace. More precisely, we assume
\begin{equation}
    \Delta,\, g \ll \omega_c \ll V,
\end{equation}
which justifies adiabatic elimination of the cavity mode. Neglecting cavity losses, the leading-order effect of this elimination results in an effective long-range interaction,
\begin{equation}\label{eq:H_xx}
    H_J = -\frac{J}{L} \sum_{i,j} \tilde{\sigma}^x_i \tilde{\sigma}^x_j,
\end{equation}
where $J = g^2 / 2\omega_c$ is the effective coupling. Unless stated otherwise, we express all energies in units of $J$. The resulting effective Hamiltonian is then
\begin{equation}
\label{eq:H_PXP2}
    \tilde{H} = -\frac{1}{L} \sum_{i,j} \tilde{\sigma}^x_i \tilde{\sigma}^x_j + \Delta \sum_i \sigma^z_i.
\end{equation}
The derivation of the effective Hamiltonian together with the next-to-leading order terms is provided in Appendix~\ref{app:H_eff}. In the remainder of this paper, we work with $\tilde{H}$ and restrict the Hilbert space to states obeying the blockade constraint. Before going into the details of the equilibrium and non-equilibrium properties of $\tilde{H}$, we give an overview of other known spin models with close connection to our system. For convenience, we call the first term in~\eqref{eq:H_PXP2} the `interaction' term in the reminder of the paper, while remembering that even in the absence of this term the system is still interacting due to the blockade condition.

The Hamiltonian $\tilde{H}$ bears a strong resemblance to the long-range transverse-field Ising model, also known as the Lipkin-Meshkov-Glick (LMG) model~\cite{LIPKIN1965188}, defined by
\begin{equation}\label{eq:H_LMG}
    H_\mathrm{LMG} = -\frac{1}{L} (S^x)^2 + \Delta\, S^z,
\end{equation}
where $S^\alpha = \sum_i \sigma^\alpha_i$ are collective spin operators. Owing to its permutation symmetry and mean-field solvability, the LMG model has been extensively studied. Its ground state exhibits a second-order phase transition from a paramagnetic to a ferromagnetic phase at $\Delta_c = 2$, in the mean-field Ising universality class~\cite{LIPKIN1965188,Ribeiro_LMG2007}. Out of equilibrium, the LMG model has also been investigated in the context of quantum quenches~\cite{Das_LMGquench2006,Marino_Eckstein_Foster_Rey_2022,rodriguez2022far} and Floquet dynamics~\cite{Russomanno_LMG2017,Lerose_Kapitza2019}. Despite its superficial similarity to the LMG model, the presence of projectors in $\tilde{H}$ imposes nontrivial kinetic constraints that fundamentally modify the system's behavior, both at equilibrium and out of equilibrium, as we demonstrate in the following sections.

Another model relevant to our discussion is the PXP model~\cite{Turner_scars2018,Ho_scars2019}, given by the following Hamiltonian
\begin{equation}
    H_\mathrm{PXP} = \sum_i \tilde{\sigma}^x_i,
\end{equation}
where the projected operator $\tilde{\sigma}^x_i$ is defined in Eq.~\eqref{eq:sx_project}. The PXP model has attracted significant attention due to its nontrivial relaxation dynamics. As first observed experimentally in Ref.~\cite{Bernien_Rydberg2017}, for certain initial states, such as the antiferromagnetic (Néel) state, the system exhibits anomalously slow thermalization, in violation of the eigenstate thermalization hypothesis (ETH). This behavior has been attributed to the presence of QMBS, which are atypical, weakly entangled eigenstates and are reminiscent of scar states in classically chaotic systems such as billiards~\cite{Serbyn_scars2021}. Initial states with large overlap with QMBS display slow entanglement growth, persistent memory of initial conditions, and long-lived coherent oscillations, in stark contrast with typical thermalizing dynamics. The connection between $\tilde{H}$ and $H_\mathrm{PXP}$ becomes evident upon noting that
\begin{equation}\label{eq:PXP2_to_PXP}
    \tilde{H} = -\frac{1}{L} \big(H_\mathrm{PXP}\big)^2 + \Delta \sum_i \sigma^z_i.
\end{equation}
This relation suggests a close correspondence between the dynamics of the two models. As we will show in Section~\ref{sec:dynamics}, the dynamics under $\tilde{H}$ is influenced by QMBS, particularly in the regime $|\Delta| \ll 1$. However, the long-range nature of the interactions in $\tilde{H}$, along with its distinct symmetries compared to $H_\mathrm{PXP}$, leads to qualitatively different dynamical behavior, most notably in the growth of entanglement and correlations.

In the following section, we analyze the ground state properties of our model, with particular emphasis on the different symmetry-breaking phases and the spectra of low-energy excitations within each phase. We then turn to the study of the model's non-equilibrium dynamics in Section~\ref{sec:dynamics}.

\section{Equilibrium Properties}\label{sec:equilibrium}
    We begin by analyzing the ground state properties of $\tilde{H}$ in Eq.~\eqref{eq:H_PXP2}. In Section~\ref{sec:groundstate}, we examine the different ground states and their associated symmetries. This is followed by a detailed study of low-energy excitations in Section~\ref{sec:excitations}. In particular, we highlight how the interplay between long-range and short-range interactions gives rise to distinctive features in both the ground state and its excitations, going beyond the paradigms established for systems with purely short-range or purely long-range interactions~\cite{Sachdev_2011,defenu_rmp2023}. We conclude our discussion of the system's equilibrium properties in Section~\ref{sec:field_theory}, where we provide an analytical treatment of the paramagnet-to-ferromagnet phase transition and the behavior of low-lying excitations near the critical point, with particular emphasis on the role of the Rydberg blockade in shaping the transition.

    \subsection{Ground State Phases}\label{sec:groundstate}

    \subsubsection{Qualitative analysis}
    Considerable insight into the ground state structure of the system as a function of $\Delta$ can be obtained on physical grounds, as we now discuss.

    We begin with the limit $\Delta \gg 1$, where the interaction term becomes negligible. In this case, the ground state is a classical paramagnet (PM), given by a product state of atomic ground states:
    \begin{equation}\label{eq:GS_param}
        \ket{\Omega_\mathrm{PM}} = \ket{0} \equiv \ket{0}_1 \ket{0}_2 \dots \ket{0}_{L-1} \ket{0}_L,
    \end{equation}
    which satisfies the blockade constraint (Eq.~\eqref{eq:blocakde}). For finite values of $\Delta$, the interaction term induces global pairwise spin flips, leading to quantum corrections to Eq.~\eqref{eq:GS_param}. As we demonstrate both analytically and numerically later, these fluctuations give rise to excitations with distinct dispersive properties, both globally and locally, due to the interplay between short-range and long-range interactions. Nevertheless, for sufficiently large $\Delta$, Eq.~\eqref{eq:GS_param} remains a good approximation to the ground state.
    
    In the opposite extreme limit $-V \ll \Delta  \ll -1$, the system favors the maximal number of excitations consistent with the Rydberg blockade. The resulting ground states are two-fold degenerate:
    \begin{equation}\label{eq:GS_neel}
        \ket{\Omega_e} = \ket{Z_2}, \quad \ket{\Omega_o} = T_1 \ket{Z_2},
    \end{equation}
    where $\ket{Z_2}$ is the N\'{e}el state,
    \begin{equation}\label{eq_neel}
        \ket{Z_2} \equiv \ket{1}_1 \ket{0}_2 \dots \ket{1}_{L-1} \ket{0}_L,
    \end{equation}
    and $T_1$ denotes translation by one lattice site. Similar results have been reported before for the PXP model~\cite{Fendley_CDW2004,Daniel_scars2023}. For large negative $\Delta$, the true ground state remains close to Eq.~\eqref{eq:GS_neel}, with only weak corrections due to interactions. These corrections are generally smaller than in the PM case, since the blockade constraint more strongly restricts quantum fluctuations around $\ket{Z_2}$ than around $\ket{0}$.

        \begin{figure}[!t]
        \centering
        \subfloat[\label{fig:ent_vs_D}]{\includegraphics[height=0.48\linewidth]{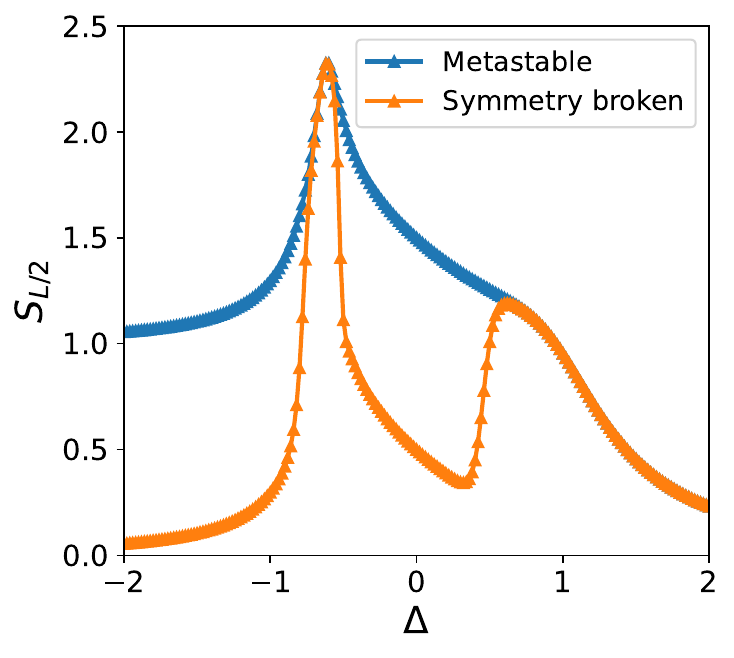}}\subfloat[\label{fig:order_param}]{\includegraphics[height=0.48\linewidth]{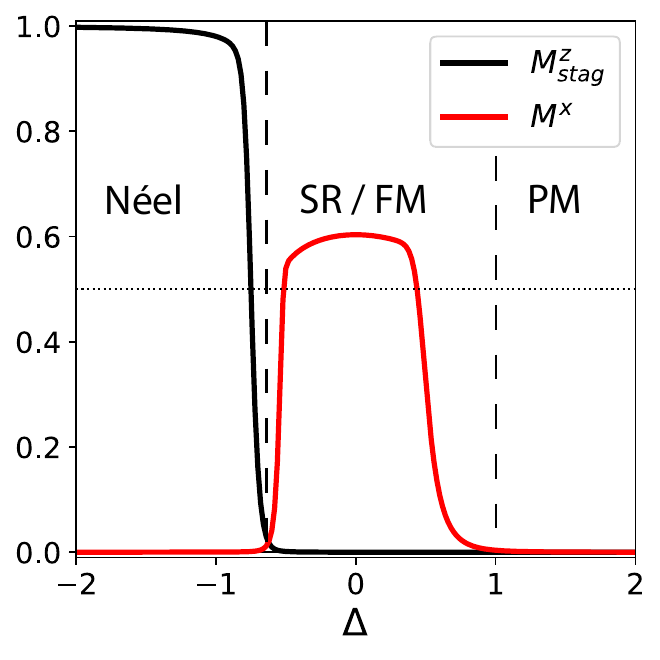}}
        \caption{\textbf{Ground state behavior of the (PXP)$^2$ model.}  (a) Ground state half-cut entanglement entropy, in the absence (blue) and presence (orange) of small symmetry breaking fields.  The peaks in the latter show phase transition points. (b) Staggered magnetization (black) and uniform magnetization (red), showing three different phases depending on the value of the field $\Delta$. Dashes show the maximum magnetization possible for a classical state in presence of blockade. Data have been obtained for a chain of 20 spins.}
    \end{figure}
    
    We now turn to the intermediate regime $|\Delta| \lesssim 1$, where interactions dominate. At $\Delta = 0$, as discussed previously (Eq.~\eqref{eq:PXP2_to_PXP}), the eigenstates of $\tilde{H}$ match those of the PXP model, which describes a constrained spin chain in a transverse field. The ground state of the PXP model exhibits finite magnetization $\langle \sigma^x_i \rangle \neq 0$~\cite{Lesanovsky_Rydberg2011}, implying the same holds for $\tilde{H}$ at $\Delta = 0$. However, in contrast to the PXP model, both its ground state and highest excited state become degenerate ground states of $\tilde{H}$ in this limit. This signals a ferromagnetic (FM) phase characterized by spontaneous breaking of the spin $Z_2$ symmetry, $\sigma^x_i \to -\sigma^x_i$. In the cavity QED context, this FM phase corresponds to a superradiant (SR) phase, where the cavity field acquires a non-zero expectation value:
    \begin{equation}
    	\langle a\rangle = \sqrt{L}\, g \, \langle \tilde{\sigma}^x \rangle.
    \end{equation}
   Finding an appropriate mean-field ansatz for the FM ground state is complicated by the presence of the blockade constraint. The simplest state satisfying the constraint while breaking both $Z_2$ and translational symmetry is
    \begin{equation}\label{eq:GS_FM_wrong}
         \ket{\rightarrow}_1 \ket{0}_2 \dots \ket{\rightarrow}_{L-1} \ket{0}_L,
    \end{equation}
    with $\ket{\rightarrow} = (\ket{0} + \ket{1}) / \sqrt{2}$. This suggests a transition from the FM to the N\'{e}el phase is associated with the restoration of the $Z_2$ symmetry. However, as we will show using exact numerical results in Section~\ref{sec:GS_numerics}, this ansatz does not describe the true FM ground state. Although the FM phase does break the $Z_2$ symmetry, it remains translationally invariant. Taking this observation into account, a more suitable ansatz for the FM groundstate is the (not normalized) projected fully polarized state given by~\cite{Moessner_slowHoles2000}
    \begin{equation}\label{eq:GS_FM_better}
        \ket{\widetilde{\Rightarrow}}\approx P_\mathrm{block}\left(\ket{\to}_1\ket{\to}_2\dots\ket{\to}_L\right),
    \end{equation}
    where $P_\mathrm{block}$ is the projector to the Rydberg blockade subspace:
    \begin{equation}
        P_\mathrm{block}= \prod_{i}(1-n_i n_{i+1}).
    \end{equation}
   The state $\ket{\widetilde{\Rightarrow}}$ breaks the spin $Z_2$ symmetry while being translationally invariant. We will compare the properties of this state with exact numerical results in the next section, showing an improved agreement compared to Eq.~\eqref{eq:GS_FM_wrong}. Last, we note that the simultaneous restoration of the spin $Z_2$ symmetry and breaking of translation symmetry at a common critical value of $\Delta$ indicates a first-order phase transition between the FM and N\'{e}el phases~\cite{Puel_Macri_2024}.

    \subsubsection{Numerical results}\label{sec:GS_numerics}

    Following the discussion of the previous section, we numerically diagonalize $\tilde{H}$ in the blockade subspace to study its ground state phases.
    
    As an unbiased indicator of phase transitions, we use the half-chain entanglement entropy $S_{L/2}$. Deep within conventional ordered phases, entanglement is typically weak, as the ground states can often be approximated by tensor product states with local structure. Near phase transitions, long-range correlations become enhanced and lead to increased entanglement. The ground state entanglement entropy of the (PXP)$^2$ Hamiltonian (Eq.~\eqref{eq:H_PXP2}) is shown in blue in Fig.~\ref{fig:ent_vs_D}. Consistent with the arguments of the previous section, the entanglement vanishes for $\Delta \gg 1$, where the system is a trivial paramagnet (PM), as described by Eq.~\eqref{eq:GS_param}. In the opposite limit, $\Delta \ll -1$, $S_{L/2}$ saturates to one. Since the logarithms are computed in base 2, this value indicates a GHZ state resulting from the symmetrization of the N\'{e}el states in Eq.~\eqref{eq:GS_neel}, i.e., $\ket{\Omega} = (\ket{Z_2} + T_1 \ket{Z_2})/\sqrt{2}$. In the thermodynamic limit, infinitesimal symmetry-breaking fields are expected to drive the system toward one of the symmetry-broken N\'eel states. Nonetheless, in finite-sized systems, such a symmetric metastable state can be adiabatically prepared~\cite{Omran_catState2019,senoo2025high}, similar to the phenomenon of supercooling in liquids~\cite{frank1952supercooling,Angell1982}. Finally, in the intermediate regime $|\Delta| \lesssim 1$, corresponding to the ferromagnetic (FM) phase, the entanglement entropy increases. A distinct peak appears near $\Delta \approx -1$, marking the transition between the N\'eel and FM phases.
    
        \begin{figure*}[!t]
    	\centering
    	\includegraphics[width=0.95\linewidth]{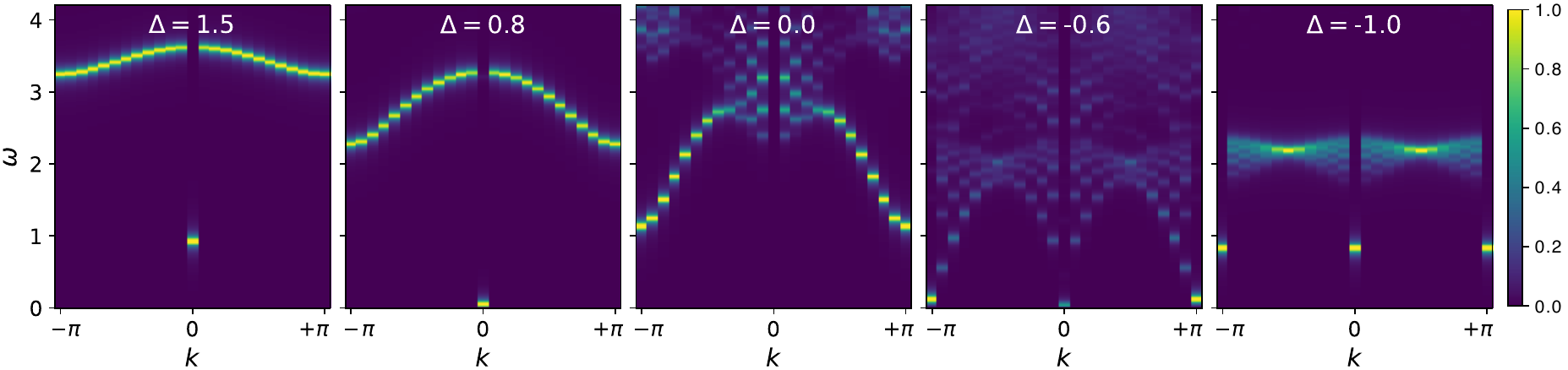}
    	\caption{Spectrum of the low-energy excitations for different values of $\Delta$, for a chain of 24 spins. The discontinuity at $k=0$ is due to the long-range cavity-induced interaction, with a vanishing gap at $\Delta \approx 0.8$, signaling a second-order transition to a FM ground state. The $k=\pi$ mode becomes gapless at $\Delta\approx -0.6$, where the ground state transforms to a N\'{e}el ordered state.}
    	\label{fig:dispersion}
    \end{figure*}
    
    To resolve the phase boundaries more sharply, we introduce a small symmetry-breaking field given by the perturbation
    \begin{equation}
        \delta H = -\epsilon \sum_i \left( \tilde{\sigma}^x_i + (-1)^i \sigma^z_i \right),
    \end{equation}
    with $\epsilon = 10^{-4}$. The entanglement entropy of the ground state of $\tilde{H} + \delta H$ is shown in orange in Fig.~\ref{fig:ent_vs_D}. In the PM phase, the result agrees with the symmetric case. In the deep N\'eel phase ($\Delta \ll -1$), the entanglement vanishes, reflecting the classical nature of the symmetry-broken $\ket{Z_2}$ state. More importantly, two distinct peaks in $S_{L/2}$ appear near $\Delta \approx \pm 1$, corresponding to the boundaries between the PM, FM, and N\'eel phases. Between these peaks, entanglement remains finite even at $\Delta = 0$, indicating that the product-state ansatz in Eq.~\eqref{eq:GS_FM_wrong} is not a valid approximation. As we elaborate below, the FM ground state and its entanglement profile are crucially affected by quantum fluctuations.
    
    To further characterize the ground state phases, we compute the order parameters associated with the expected symmetry-broken phases as functions of $\Delta$. Specifically, we evaluate the average magnetization,
    \begin{equation}
        M^x = \frac{1}{L} \sum_i \langle \sigma^x_i\rangle,
    \end{equation}
    which serves as the FM order parameter, and the staggered magnetization,
    \begin{equation}
        M^z_\mathrm{stag} = \frac{1}{L}  \sum_i (-1)^i \langle\sigma^z_i \rangle,
    \end{equation}
    which acts as the N\'eel order parameter. As shown in Fig.~\ref{fig:order_param}, the phase boundaries inferred from these order parameters align with those obtained from the entanglement entropy. The behavior in both the PM and N\'eel phases is consistent with classical ground states given by Eqs.~\eqref{eq:GS_param} and~\eqref{eq:GS_neel}, respectively. In particular, $M^z_\mathrm{stag}$ reaches its maximal value quickly for $\Delta \lesssim -1$.
    
    A striking feature of the FM phase is the absence of staggered magnetization, i.e., no translation symmetry breaking, despite the presence of strong Rydberg blockade. This further confirms that the ansatz in Eq.~\eqref{eq:GS_FM_wrong}, which predicts finite $M^z_\mathrm{stag}$, does not accurately describe the FM ground state. Furthermore, the maximum value of $M^x$, observed at $\Delta = 0$, exceeds the maximum possible value attainable by any classical state that respects the blockade constraint:
    \begin{equation}
        \max(M^x) \approx 0.6 > \max(M^x)_\mathrm{cl} = 0.5.
    \end{equation}
    This excess highlights the presence of quantum fluctuations in the FM phase due strong blockade. In Section~\ref{sec:groundstate}, we also provided a translationally invariant ansatz for the FM groundstate in Eq.~\eqref{eq:GS_FM_better}. Calculation of the average magnetization for this state yields:
    \begin{equation}
        \frac{\bra{\widetilde{\Rightarrow}} \sigma^x_i \ket{\widetilde{\Rightarrow}}}{\braket{\widetilde{\Rightarrow}}}= \frac{2}{2+\varphi}+O(\varphi^{-2L})\approx 0.55,
    \end{equation}
    where $\varphi=(1+\sqrt{5})/2$ is the golden ratio. We see that the magnetization of $\ket{\widetilde{\Rightarrow}}$ exceeds the maximum classical limit, although it is still smaller than the exact value. 
    Despite this difference, $\ket{\widetilde{\Rightarrow}}$ provides a good approximation of the groundstate at $\Delta=0$.
    
    Based on our numerical results, we conclude that the PM–FM transition is a second-order phase transition associated with the spontaneous breaking of the spin $Z_2$ symmetry, similar to the Dicke phase transition. In this regime, low-energy physics is dominated by long-range interactions, while the blockade modifies the structure of gapped excitations, as discussed in Section~\ref{sec:excitations}. In contrast, the FM–N\'eel transition connects two distinct symmetry-broken phases and is characterized by a sharp change in both order parameters, as seen in Fig.~\ref{fig:order_param}. Our findings are consistent with those in Refs.~\cite{Puel_Macri_2024,Bacciconi_Xavier_Marinelli_Bhakuni_Dalmonte_2025}, indicating a first-order transition.

    \subsection{Low energy excitations}\label{sec:excitations}
    Following the discussion of the groundstate in the previous section, we now turn to the low-energy excitations of the system above the groundstate.

    To study the low-energy modes, we use the following momentum-resolved spectral density
    \begin{equation}\label{eq:spectral_density}
        A(k,\omega)\equiv \sum_{\alpha} \abs{\bra{\alpha}\sigma^x_{k}\ket{\Omega}}^2 \,\delta(\omega - E_\alpha+E_\Omega),
    \end{equation}
    where $\ket{\alpha}$ and $E_\alpha$ are respectively the eigenstates and energies of $\tilde{H}$ with groundstate $\ket{\Omega}$. $\sigma^x_{k}=\sum_j e^{-ikj} \sigma^x_j/\sqrt{L}$ is the Fourier transform of spin operator in terms of lattice momentum $k=2\pi n_k/L$ with $-L/2<n_k\le L/2$. The dependence of $A(k,\omega)$ on $\sigma^x$ makes it particularly sensitive to the excitations of the PM phase and its transition to the FM state.

    We have displayed the normalized spectral density for different values of $\Delta$ in Fig.~\ref{fig:dispersion}. Starting from the PM phase ($\Delta=1.5$), the spectrum has two notable features. First, is the presence of a continuum of dispersive modes at finite momenta, corresponding to magnons, separated by a large energy gap from the groundstate. Second, is the single mid-gap zero-momentum mode, which is a signature of long-range interactions in the system, since, in the absence of Rydberg blockade, finite-momentum excitations are degenerate and dispersionless~\cite{defenu_rmp2023}. The blockade-induced screening of spin operators by projectors in $\tilde{H}$ lifts the degeneracy of the finite-momentum manifold. As a result, excitations at momenta close to $\boldsymbol{k}=\pi$ can reduce their energies, which demonstrates the tendency of the system to form antiferromagnetic correlations, consistent with Rydberg interactions. We note that the dispersion of excitations is sensitive to the system size, with the excitation bandwidth decreasing as $L$ increases. We analytically explain this numerical observation later in Section~\ref{sec:field_theory}. This suggests that in the thermodynamic limit, the effects of the Rydberg blockade become negligible in the PM phase and near the PM–FM transition. However, in experimentally relevant regimes, where only a few tens of atoms can be trapped inside the cavity using optical tweezers, dispersive modes persist and play a crucial role in interpreting experimental observations.

    By approaching the PM to FM transition, the finite-momentum modes become more dispersive, with the $\boldsymbol{k}=\pi$ mode reducing its energy gap. Similarly, zero-momentum mode becomes softer, and its energy gap vanishes faster as we decrease $\Delta$. As a result, the phase transition corresponds to the condensation of zero-momentum mode, and the formation of FM order. The blockade merely causes enhanced quantum fluctuations in the FM phase. Decreasing $\Delta$ further enhances dispersive modes until we reach the FM to N\'{e}el transition point. At this point, the excitation gap at $\boldsymbol{k}=\pi$ vanishes, while the zero-momentum mode acquires an energy gap. The vanishing of the energy gap on both sides of the transition is not a typical feature of first order transitions, and could be a consequence of the kinetically constrained nature of the system.  Deep in the N\'{e}el phase, finite-momentum modes become weakly dispersive, while being significantly broadened.

    \subsection{Field theoretical treatment}\label{sec:field_theory}

    We develop an analytical framework that provides insight into several numerical observations reported in previous sections. Our approach, inspired by methods used to study quantum Ising transitions~\cite{Sachdev_2011,Ye_rotorglass93,strack_dickeglass2011}, involves mapping the original spin degrees of freedom to soft-spin variables via the replacement
\begin{equation}
    \sigma^x_i \to \varphi_i.
\end{equation}

To begin, we consider the non-interacting part of the Hamiltonian. In the absence of the blockade constraint, this mapping yields
\begin{equation}\label{eq:Hz_to_Hphi}
    \Delta \sum_i \sigma^z_i \to r \sum_i \left( \pi_i^2 + \varphi_i^2 \right),
\end{equation}
where $\pi_i$ is the canonical conjugate momentum to $\varphi_i$, satisfying $[\varphi_i, \pi_j] = i\delta_{ij}$. The parameter $r$ should be chosen such that the excitation energy in this effective model matches that of the original spin model deep in the PM phase.

Since $(\sigma^x)^2 = 1$, we must impose a constraint on the magnitude of $\varphi_i$. This is most naturally done in the path integral formalism, where the imaginary-time action becomes
\begin{equation}
    S_0 = \sum_i \int d\tau\, \left[ \frac{1}{4r} (\partial_\tau \varphi_i)^2 + r \varphi_i^2 - \chi_i(\tau)(\varphi_i^2 - 1) \right],
\end{equation}
with $\chi_i(\tau)$ a time-dependent Lagrange multiplier field enforcing the unit-length constraint $\varphi_i^2 = 1$.

To incorporate the Rydberg blockade, which forbids simultaneous excitation of neighboring sites, we introduce a second Lagrange multiplier field $\lambda_i(\tau)$ to impose the constraint $\varphi_i \varphi_{i+1} = 0$:
\begin{equation}
    S_\mathrm{bl} = -\sum_i \int d\tau\, \lambda_i(\tau) \, \varphi_i \varphi_{i+1} .
\end{equation}

The final component is the long-range interaction. Using Eq.~\eqref{eq:H_xx}, its contribution to the action takes the form
\begin{equation}
    S_\mathrm{int} = -\frac{J}{L} \sum_{i,j} \int d\tau\, \varphi_i \varphi_j.
\end{equation}

   \begin{figure}[!t]
        \centering
        \includegraphics[width=0.6\linewidth]{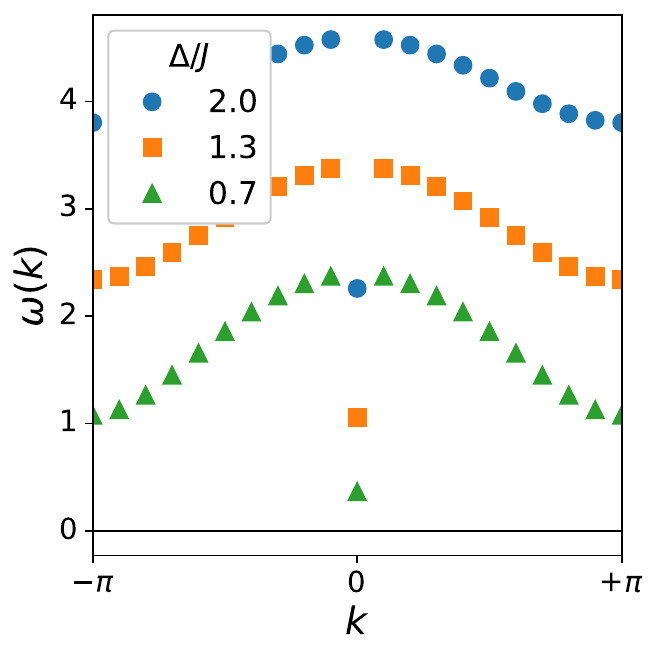}
        \caption{Dispersion of excitations in the paramagnetic phase and close to the FM transition, obtained from the soft-spin approximation for a chain of 24 spins.}
        \label{fig:disper_analytic}
    \end{figure}

     The model is thus described by the sum of the three contributions introduced above:
\begin{equation}
    S = S_0 + S_\mathrm{bl} + S_\mathrm{int}.
\end{equation}
    We focus on the paramagnetic (PM) regime, where $\langle \varphi_i \rangle$ vanishes. To proceed analytically, we make a simplifying approximation by treating the Lagrange multiplier fields as uniform and time-independent parameters:
    \begin{equation}
        \chi_i(\tau) = \chi, \quad \lambda_i(\tau) = \lambda,
    \end{equation}
    which allows the action to be written in momentum and frequency space as
    \begin{equation}
        S = \frac{1}{4r}\sum_k \int_0^\infty \frac{d\omega}{2\pi} \left( \omega^2 + \omega_k^2 \right) \abs{\varphi(k,\omega)}^2,
    \end{equation}
    where the excitation dispersion is given by
    \begin{equation}
        \omega_k = 2 \sqrt{r \left( r - \chi - \lambda \cos{k} - J \delta_{k,0} \right)}.
    \end{equation}

\begin{figure*}[!t]
		\centering
		\subfloat[\label{fig:overlap}]{\includegraphics[height=0.23\linewidth]{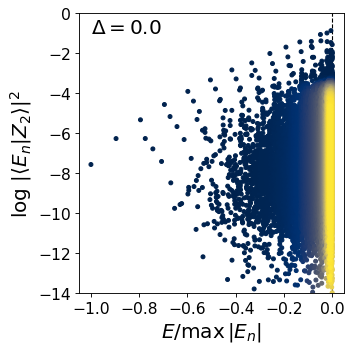}}\subfloat[\label{fig:overlap_finite_Delta}]{\includegraphics[height=0.23\linewidth]{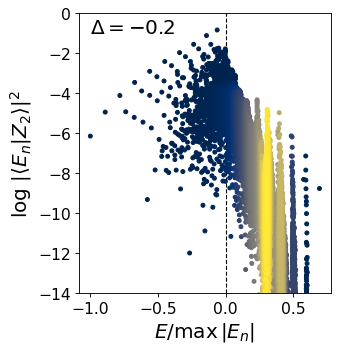}}\subfloat[\label{fig:levelstats}]{\includegraphics[height=0.23\linewidth]{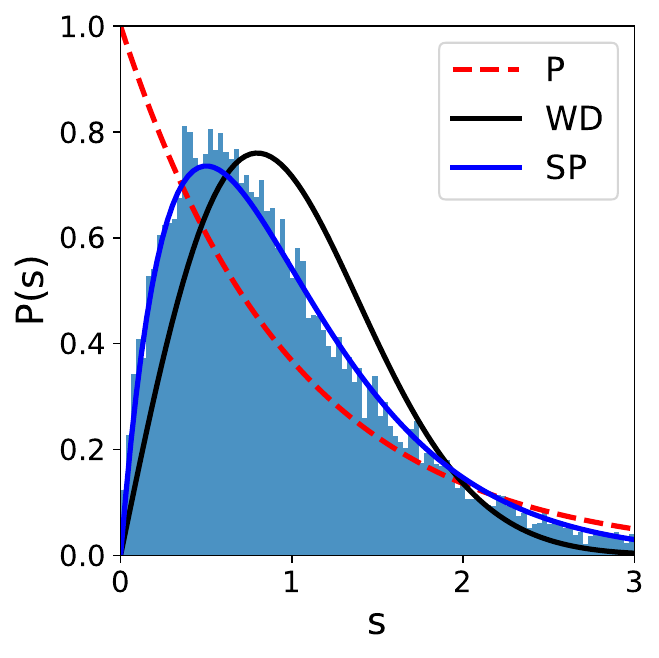}}\subfloat[ \label{fig:Czz_vs_t_Z2}]{\includegraphics[height=0.23\linewidth]{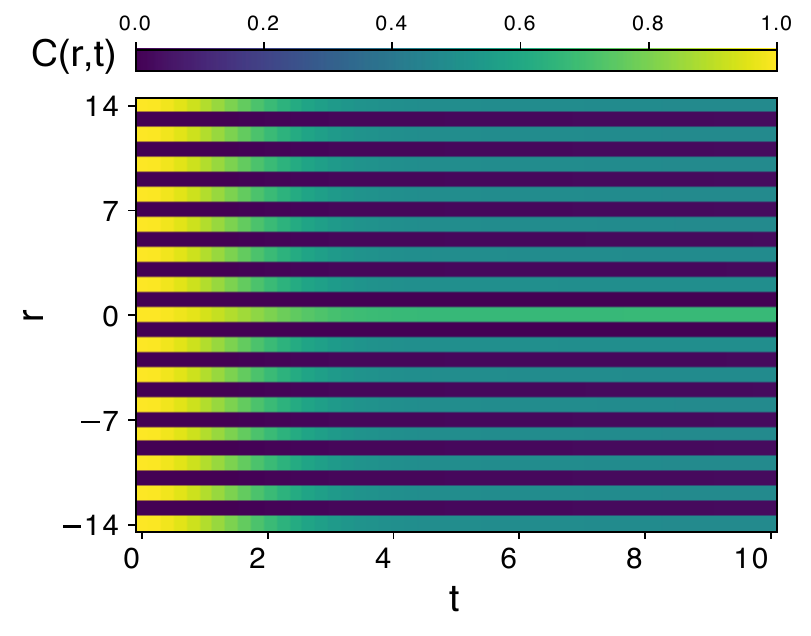}}
		\caption{(a) and (b), the overlap between the N\'{e}el state and the eigenstates of the Hamiltonian for $\Delta=0$ and $\Delta=-0.2$, showing the presence of scars in the spectrum. The colormap indicates the density of the points. (c) Energy level statistics in the zero-momentum and inversion-symmetric subspace, compared to Poisson, Wigner-Dyson, and semi-Poisson distributions. (d) Correlation dynamics for $\ket{\psi_0}=\ket{Z_2}$. The long-time persistence of staggered magnetization for $\Delta=0$ is due to the overlap of the initial state with scars. $L=30$ for (a)-(c) and $28$ for (d).}
		
	\end{figure*}
    
    We note that the long-range interaction modifies the dispersion at zero momentum, by introducing a discontinuity through the $J \delta_{k,0}$ term. The dispersion of finite-momentum modes arises due to the interplay of the blockade and long-range interactions, as encoded by $\lambda$. The values of $\chi$ and $\lambda$ are determined by enforcing the constraints at the level of expectation values. The unit-length constraint leads to
    \begin{equation}\label{eq:phi_unit}
        \langle \varphi_i^2 \rangle = 1 \quad \Rightarrow \quad \frac{1}{L} \sum_k \frac{r}{\omega_k} = 1,
    \end{equation}
    while the Rydberg blockade constraint imposes
    \begin{equation}\label{eq:phi_blockade}
        \langle \varphi_i \varphi_{i+1} \rangle = 0 \quad \Rightarrow \quad \frac{1}{L} \sum_k \frac{\cos{k}}{\omega_k} = 0.
    \end{equation}
    
    Eqs.~\eqref{eq:phi_unit} and~\eqref{eq:phi_blockade} can be solved in the large-$L$ limit. Moreover, to match the excitation gap with that of the original model in the deep PM regime, we require consistency with the microscopic energy scale. The result is
    \begin{equation}\label{eq:lambda_vs_J}
        r = 2\Delta, \quad \chi \approx \frac{3}{2} \Delta, \quad \lambda \approx -\frac{2\Delta}{L} \left( \sqrt{\frac{\Delta}{\Delta - 2J}} - 1 \right),
    \end{equation}
    to leading order in $1/L$. Eq.~\eqref{eq:lambda_vs_J} is consistent with the numerical results presented in Section~\ref{sec:excitations}, which attributed the finite bandwidth of excitations to the Rydberg blockade. In particular, the bandwidth increases as $\Delta$ approaches the phase transition, and scales inversely with the system size, in agreement with Eq.~\eqref{eq:lambda_vs_J}.
    
    We emphasize that this result relies on the smallness of $\lambda$, which holds due to its $1/L$ scaling. However, as $\Delta \to 2J$, this approximation breaks down, and Eqs.~\eqref{eq:phi_unit} and~\eqref{eq:phi_blockade} must be solved numerically. The dispersion $\omega_k$ obtained from the analytical approach is shown in Fig.~\ref{fig:disper_analytic}, and it agrees qualitatively with the exact numerical results shown in Fig.~\ref{fig:dispersion} for the PM regime.

    \section{Non-equilibrium dynamics}\label{sec:dynamics}

    We now turn to the dynamics of the system far from equilibrium. Due to the Rydberg blockade, the model is subject to kinetic constraints. Such constraints are known to give rise to unconventional dynamics, including persistent macroscopic oscillations and slow thermalization. In addition, the presence of long-range interactions leads to qualitatively distinct behavior compared to systems with purely local interactions, such as enhanced collective effects. The model defined in Eq.~\eqref{eq:H_PXP2} combines both of these ingredients, and we demonstrate below how their interplay shapes the non-equilibrium dynamics.

    In the following, we first investigate the scarred spectrum of the system in Section~\ref{sec:scars}, showing its close connection to the scars of the PXP model. We then address entanglement dynamics in Section~\ref{sec:ent_dynamics}, showing that despite the similarities of the two models, they display qualitatively different entanglement growth.

    \subsection{Quantum many-body scars}\label{sec:scars}

    The presence of QMBS in the energy spectrum of the system is evident from its connection to the PXP model, as expressed in Eq.~\eqref{eq:PXP2_to_PXP}. Similar to the PXP model~\cite{Turner_scars2018}, density-wave states such as the N\'eel state defined in Eq.~\eqref{eq_neel} have significant overlap with scarred eigenstates. In Fig.~\ref{fig:overlap}, we show the overlap between the N\'eel state and the eigenstates of $\tilde{H}$ at $\Delta = 0$, clearly revealing the presence of scars in the spectrum. However, unlike in the PXP model, the level spacing between scarred states is no longer approximately constant. This significant overlap with QMBS explains the absence of thermalization when the system is initialized in the state $\ket{Z_2}$.
    
    We also analyze the level statistics of the system. In the case of the LMG model (Eq.~\eqref{eq:H_LMG}), the spectrum exhibits non-generic level statistics that does not correspond to those of chaotic or integrable systems~\cite{LIPKIN1965188,defenu_rmp2023}. Since our model shares structural features with that case, one might expect its spectrum to inherit similar traits. However, as in the PXP model, the unfolded distribution of level spacings exhibits fully chaotic behavior, as shown in Fig.~\ref{fig:levelstats}  for zero-momentum and inversion-symmetric energy eigenstates. In fact, at $\Delta = 0$, the level statistics of $\tilde{H}$ matches those of the PXP model, which agrees well with a semi-Poisson distribution~\cite{Turner_scars2018}
    \begin{equation}
        P(s) = 4s e^{-2s}.
    \end{equation}
    This can be demonstrated by analyzing the unfolded level spacings,
    \begin{equation}
        s_i \equiv (\epsilon_{i+1} - \epsilon_i) \left.\dv{\mathcal{N}}{\epsilon}\right|_{\epsilon_i},
    \end{equation}
    where $\epsilon_i$ are the energy levels and $\mathcal{N}(\epsilon)$ is a smooth interpolation of the level counting function. According to Eq.~\eqref{eq:PXP2_to_PXP}, for $\Delta = 0$, we have $\epsilon \sim \epsilon_\mathrm{PXP}^2$, implying that
    \begin{equation}
        s_i = (s_i)_{\mathrm{PXP}},
    \end{equation}
    and hence the level spacing distributions of the two models coincide.

    Thus far, we have focused on the case $\Delta = 0$, where a one-to-one mapping to the PXP model exists (Eq.~\eqref{eq:PXP2_to_PXP}). However, one can also study the stability of quantum many-body scars at finite values of $\Delta$. As shown in Fig.~\ref{fig:overlap_finite_Delta}, scarred eigenstates persist over a finite range of $\Delta$. For large positive or negative values of $\Delta$, these states gradually evolve into simple product eigenstates of $\tilde{H}$, consistent with the asymptotic limits $\Delta \to \pm \infty$.

	In order to demonstrate the dynamical consequence of QMBS, we consider the evolution of the two-point correlation function
	\begin{equation}
		C(r,t) = \langle n_i(t)\, n_{i+r}(t) \rangle,
	\end{equation}
	after initializing the system in the N\'{e}el state. In case of ergodic dynamics, we expect the initial density-wave profile to quickly relax to a high temperature thermal state with uniform excitation density. A signature of scarred dynamics and weak ergodicity breaking is the persistence of initial density-wave profile for long times. Starting from a N\'eel state and for $\Delta=0$, as shown in Fig.~\ref{fig:Czz_vs_t_Z2}, we see that the initial staggered pattern persists for long times, as a consequence of the significant overlap between $\ket{Z_2}$ and the scarred eigenstates of $\tilde{H}$. However, in contrast to the PXP model, $C(r,t)$ does not display persistent coherent oscillations. This distinction can be attributed to the absence of approximate commensurability among the scarred level spacings in $\tilde{H}$, which inhibits the formation of regular revival dynamics typically observed in QMBS associated with the PXP model~\cite{Turner_scars2018}.

    Another hallmark of scarred dynamics and ergodicity breaking is the slow growth of entanglement. In our model, this behavior reflects an intriguing interplay between the kinetic constraint imposed by the Rydberg blockade and the long-range nature of the interactions, as we discuss in detail in the following section.

    \subsection{Entanglement Dynamics}\label{sec:ent_dynamics}

    Entanglement is a defining feature of quantum systems that, in simple terms, quantifies how much information about the rest of the system is imprinted on a given subsystem. Following a sudden change in system parameters, generic (non-integrable) systems typically exhibit chaotic dynamics, characterized by a linear growth of entanglement in time. In contrast, integrable systems display slower entanglement growth due to the presence of conserved quantities. A hallmark of systems hosting QMBS, despite being chaotic in the bulk of their spectrum (see Fig.~\ref{fig:levelstats}), is the unusually slow growth of entanglement when the initial state has significant overlap with scarred eigenstates~\cite{Turner_scars2018,Serbyn_scars2021}. Interestingly, slow entanglement growth is also found in certain non-integrable systems with long-range interactions. In such cases, entanglement can grow logarithmically in time, a behavior that may appear counterintuitive at first glance, as the non-locality of interactions would be expected to lead to the fast spreading of entanglement across the system. However, this behavior can be understood in terms of the production of collective excitations that dominate the dynamics~\cite{Lerose_SlowGrowth2020} (see also the discussion below).

    Notably, the cavity-Rydberg setup considered in this work naturally embeds both QMBS and long-range interactions within a single model. As a result, the entanglement dynamics exhibits a hybrid character, partly resembling scarred dynamics and partly reflecting features typical of long-range interacting systems, with certain aspects of the evolution admitting interpretations from both perspectives.

    In the following, we first examine entanglement dynamics at $\Delta = 0$ in Section~\ref{sec:ent_dyn_zero_D}, and then analyze the case of finite $\Delta$ in Section~\ref{sec:ent_dyn_finite_D}.

    \subsubsection{Zero detuning ($\Delta=0$)}\label{sec:ent_dyn_zero_D}

    As discussed earlier, the case $\Delta = 0$ is closely related to the standard PXP model. Consequently, we follow a similar approach to analyzing entanglement dynamics by tracking the evolution of the half-chain entanglement entropy,
    \begin{equation}
        S_{L/2}(t) = -\Tr_A \big( \rho_A(t) \ln \rho_A(t) \big),
    \end{equation}
    where $\rho_A(t)$ is the reduced density matrix of a subsystem $A$ of size $L/2$, obtained by tracing over the complementary subsystem $B$, $\rho_A(t) = \Tr_B \ketbra{\psi(t)}{\psi(t)}$, with $\ket{\psi(t)} = e^{-i \tilde{H} t} \ket{\psi_0}$ the time-evolved state. We expect distinct behaviors of $S_{L/2}(t)$ depending on whether the initial state $\ket{\psi_0}$ has significant overlap with scarred eigenstates.

\begin{figure}[!t]
        \centering
        \includegraphics[width=0.8\linewidth]{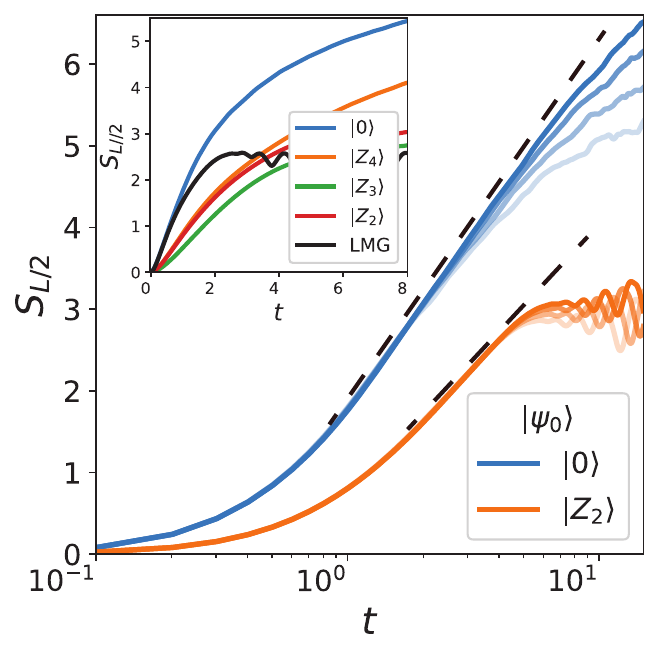}
        \caption{Evolution of entanglement entropy in the (PXP)$^2$ model at $\Delta = 0$, starting from the vacuum (blue) and N\'eel (orange) states, for system sizes $L = 20, 22, 24, 26$ (lighter to darker curves). Dashed lines indicate logarithmic growth. Inset: Entanglement dynamics for various density-wave initial states with different periodicities in the (PXP)$^2$ model, compared to the LMG model, all evaluated at $\Delta = 0$ and $L = 24$.}
        \label{fig:S_vs_logt}
    \end{figure}

    We show the time evolution of $S_{L/2}(t)$ in Fig.~\ref{fig:S_vs_logt} for two different initial states: the vacuum state $\ket{0}$ and the N\'eel state $\ket{Z_2}$, across various system sizes. The primary distinction between the two is the markedly slower entanglement growth observed for the N\'eel state, which has significant overlap with scarred eigenstates (Fig.~\ref{fig:overlap}). In this respect, the influence of scars mirrors that seen in the PXP model. It is worth noting that, unlike in the PXP model, the two initial states do not have the same energy with respect to the (PXP)$^2$ Hamiltonian. However, their energy difference is sub-extensive in the system size, scaling as $\bra{0} \tilde{H} \ket{0} - \bra{Z_2} \tilde{H} \ket{Z_2} \sim O(1)$, and is therefore negligible for system sizes that we work with.

    Despite the broadly similar influence of scars on entanglement dynamics in the PXP and (PXP)$^2$ models, the temporal growth of entanglement follows distinct trends in the two cases. In the former, entanglement entropy grows linearly in time, albeit with different rates for $\ket{0}$ and $\ket{Z_2}$~\cite{Turner_scars2018,Ho_scars2019}. In contrast, for the (PXP)$^2$ model, the growth is logarithmic, as illustrated by the dashed lines in Fig.~\ref{fig:S_vs_logt}. Logarithmic growth implies that entanglement spreads faster than linear at early times but slows down significantly at later times. This behavior is a direct consequence of long-range interactions, which are not constrained by locality and can entangle distant regions of the system almost instantaneously.

    Similar logarithmic growth has been observed in other systems with long-range interactions, where couplings decay with distance as $r^{-\alpha}$ for $\alpha < d$, with $d$ the spatial dimension~\cite{Schachenmater_EntGrowthLongRange2013,Buyskikh_EntGrowthLongrange2016,Pappalardi_EntGrowthLongrange2018}. This behavior was later explained in Ref.~\cite{Lerose_SlowGrowth2020} in terms of the linear-in-time proliferation of magnons, whose contribution to the entanglement entropy scales as
    \begin{equation}\label{eq:S_logn_logt}
        S_{L/2}(t) \sim \log n(t) \sim \log t,
    \end{equation}
    where $n(t)$ is the total magnon population. In this picture, the condition $\alpha < d$ ensures that magnon scattering is suppressed in the thermodynamic limit, allowing the free-magnon approximation to remain valid. As we will show below, and later in Section~\ref{sec:ent_dyn_finite_D} for the case $\Delta \neq 0$, this picture also provides a useful framework for interpreting various features of entanglement dynamics in our model.

    To further illustrate the connection between our model and other long-range interacting systems, we compare the entanglement dynamics of our model with that of the LMG model (Eq.~\eqref{eq:H_LMG}). In the inset of Fig.~\ref{fig:S_vs_logt}, we present the evolution of the half-cut entanglement entropy generated by the two models for several initial product states: the vacuum state, the N\'eel state, and the density-wave states $\ket{Z_3}$ and $\ket{Z_4}$, which have periodicities of three and four sites, respectively. For the LMG model, the three density-wave states exhibit identical entanglement dynamics to the vacuum state, as they are related to it by a $\pi$-rotation around the $x$-axis applied to the occupied sites, which is a symmetry of $H_\mathrm{LMG}$ for $\Delta=0$. We find that, when initialized in the vacuum state, the entanglement growth in both the (PXP)$^2$ and LMG models agrees at early times, with their corresponding curves nearly overlapping. This agreement can be attributed to the fact that, at short times, the number of spin flips remains small, and the blockade projectors (Eq.~\eqref{eq:sx_project}) have a negligible effect on the dynamics.

    \begin{figure*}[!t]
        \centering
        \subfloat[\label{fig:St_vs_pos_D}]{\includegraphics[width=0.3\linewidth]{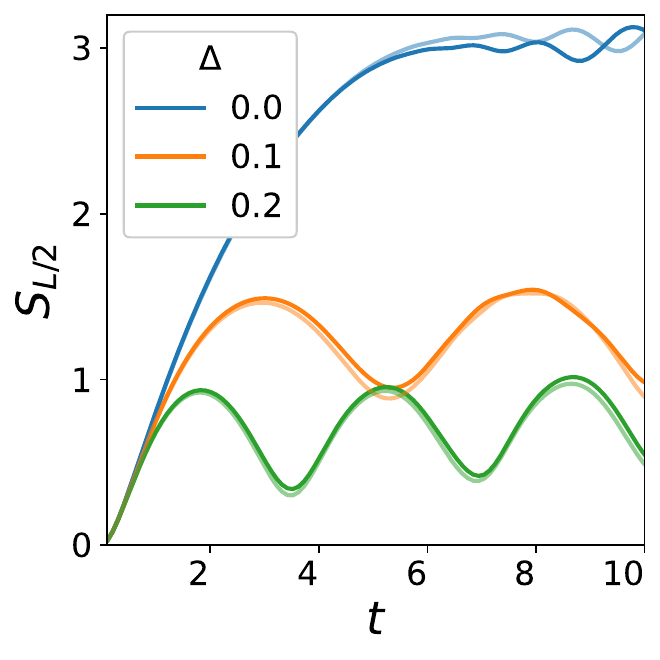}}\quad\subfloat[\label{fig:St_vs_neg_D}]{\includegraphics[width=0.3\linewidth]{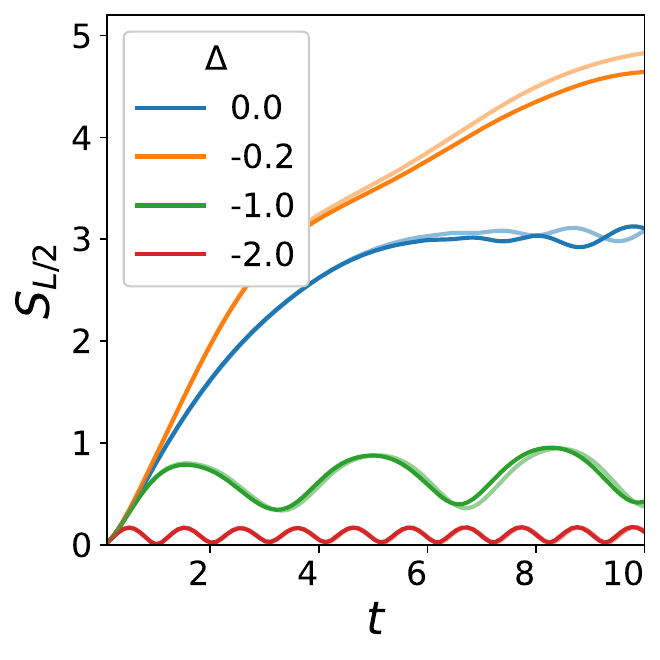}}\quad\subfloat[\label{fig:S_vs_t_deform}]{\includegraphics[width=0.3\linewidth]{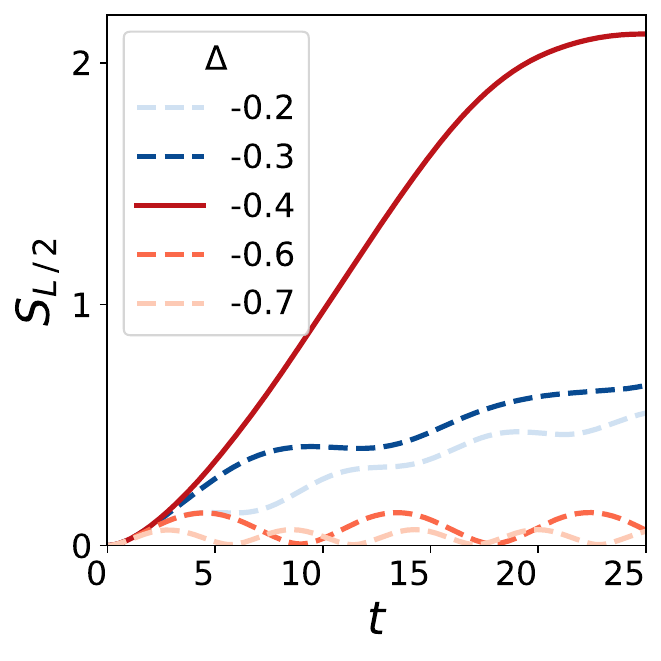}}
        \caption{Entanglement dynamics in the (PXP)$^2$ model starting from the N\'eel state and for (a) $\Delta > 0$ and (b) $\Delta < 0$, for chain sizes $L = 24$ and $26$ (light and dark curves, respectively). In panel (b), the entanglement growth rate exhibits a clear maximum within the range $-1 \lesssim \Delta < 0$. (c) Entanglement dynamics of the deformed model (Eq.~\eqref{eq:H_deformed}) with $\chi = 0.1$ and $L = 22$, displaying a peak at $\Delta \approx -0.4$, consistent with resonant excitation of magnons in the system. }
    \end{figure*}

    The above observation should be contrasted with the case where the system is initialized in the $\ket{Z_2}$ state, for which the blockade projectors impose maximal constraints on the dynamics. Remarkably, we still observe logarithmic entanglement growth in this case, despite the relevance of the blockade. This behavior can be understood as follows. We decompose the Hamiltonian $\tilde{H}$ from Eq.~\eqref{eq:H_PXP2} into three parts:
    \begin{equation}\label{eq:H_sublatt_decomp}
        \tilde{H} = \tilde{H}_\mathrm{e} + \tilde{H}_\mathrm{o} + \tilde{H}_\mathrm{eo},
    \end{equation}
    where
    \begin{equation}\label{eq:H_even}
        \tilde{H}_\eta = -\frac{1}{L} \sum_{i,j \in \eta} \tilde{\sigma}^x_i \tilde{\sigma}^x_j + \Delta \sum_{i \in \eta} \sigma^z_i, \quad (\eta = \mathrm{e, o}),
    \end{equation}
    generates spin-flip dynamics within each of the even and odd sublattices, and
    \begin{equation}\label{eq:H_even_odd}
        \tilde{H}_\mathrm{eo} = -\frac{1}{L} \sum_{i \in e} \sum_{j \in o} \tilde{\sigma}^x_i \tilde{\sigma}^x_j + \mathrm{h.c.},
    \end{equation}
    corresponds to inter-sublattice spin-flip processes. In this decomposition, $\tilde{H}_\eta$ contributes negligibly to the dynamics at early times if the corresponding sublattice $\eta$ is initially unoccupied, as is the case in the N\'eel state. Similarly, $\tilde{H}_\mathrm{eo}$ is suppressed due to the blockade constraint. The remaining non-trivial contribution arises from $\tilde{H}_\eta$ acting on the initially occupied sublattice. Since all adjacent sites on this sublattice are separated by at least one unoccupied site, the blockade constraint becomes ineffective within it, and $\tilde{H}_\eta$ can be approximately treated as an LMG model without projectors. This effective dynamics on a single sublattice then explains the logarithmic growth of entanglement entropy observed at early times.

    In the PXP model, density-wave states with longer periodicities have progressively smaller overlap with QMBS and, in the limit of large period, asymptotically behave like the vacuum state~\cite{Turner_scars2018}. A notable observation about the (PXP)$^2$ model, as shown in the inset of Fig.~\ref{fig:S_vs_logt}, is the slower entanglement growth of the $\ket{Z_3}$ state compared to $\ket{Z_2}$. Moreover, $\ket{Z_2}$ and $\ket{Z_4}$ exhibit identical growth at early times. Both of these observations can again be explained by the interplay between the Rydberg blockade and long-range interactions. The similarity between $\ket{Z_2}$ and $\ket{Z_4}$ arises from the fact that the system can be partitioned into two sublattices, and spin flips in the fully unoccupied sublattice are suppressed by the blockade constraint. As a result, the early-time dynamics can again be mapped to an LMG model restricted to the occupied sublattice, using Eqs.~\eqref{eq:H_sublatt_decomp}–\eqref{eq:H_even_odd}.For the $\ket{Z_3}$ state, the system can be partitioned into three sublattices, with only one of them initially occupied. Due to the blockade constraint, spin flips in the two unoccupied sublattices are suppressed, effectively restricting the dynamics to the occupied sublattice. Consequently, the early-time evolution can be approximately mapped to an LMG model defined over one-third of the chain, with a correspondingly reduced phase space for generating excitations. We remark that, since the overlap of $\ket{Z_3}$ with scars is smaller than that of $\ket{Z_2}$, one expects the entanglement entropy of the former to eventually surpass that of the latter at late times. However, strong finite-size-induced oscillations obscure this behavior in the observed entanglement dynamics.

    In summary, the slower entanglement growth observed for the N\'eel state can be understood from two complementary perspectives. From the viewpoint of scarred dynamics, it arises due to the significant overlap of the initial state with scarred eigenstates. In contrast, within the magnon picture, it can be attributed to the reduced phase space available for magnon creation. We emphasize, however, that the magnon-based explanation is limited to early and intermediate-time dynamics. At later times, the occupation of both sublattices becomes appreciable, and the assumptions underlying the magnon picture break down. Consequently, the long-time saturation of entanglement entropy to lower values for the N\'eel state must be attributed to the presence of scars in the spectrum.

    \subsubsection{Finite detuning ($\Delta\neq0$)}\label{sec:ent_dyn_finite_D}

    Having discussed the case of vanishing $\Delta = 0$, we now turn to the more general regime of finite $\Delta$. As we shall demonstrate, the interplay between the blockade constraint and long-range interactions gives rise to a non-trivial dependence of the entanglement dynamics on $\Delta$, exhibiting a marked asymmetry between positive and negative values.

    We begin by considering the regime $\Delta > 0$. The half-cut entanglement entropy, evaluated for time evolution from the N\'eel state, is shown in Fig.~\ref{fig:St_vs_pos_D}. As $\Delta$ increases, the entanglement growth is progressively suppressed. This behavior can be understood from two complementary perspectives. In terms of eigenstate structure, larger $\Delta$ enhances the overlap between the Hamiltonian eigenstates and those of $\sigma^z_i$, including the N\'eel state itself, thereby limiting dynamics. In the magnon picture, increasing $\Delta$ raises the magnon energy gap, which in turn suppresses spin-flip processes generated by the interaction term. Overall, the behavior observed for positive $\Delta$ is consistent with expectations.

    While one might expect a similar suppression of entanglement growth for $\Delta < 0$, the observed behavior is notably different. As shown in Fig.~\ref{fig:St_vs_neg_D}, decreasing $\Delta$ initially enhances the entanglement growth rate, which reaches a maximum around $\Delta \approx -0.5$. Upon further reduction of $\Delta$, the dynamics begins to follow the expected trend: entanglement growth becomes increasingly suppressed, reflecting the growing overlap between the N\'eel state and the ground state of $\tilde{H}$ for $\Delta \lesssim -1$. The non-monotonic behavior, and in particular the existence of a maximum, again originates from the interplay between the Rydberg blockade and long-range interactions, as we elaborate below.

    Our argument starts with the following deformation of $\tilde{H}$ (Eq.~\eqref{eq:H_PXP2}):
    \begin{equation}\label{eq:H_deformed}
        \tilde{H}(\chi) = \tilde{H}_\mathrm{XY} + \tilde{H}_\mathrm{drive}(\chi),
    \end{equation}
    where $\tilde{H}_\mathrm{XY}$ is the spin-conserving part,
    \begin{equation}
        \tilde{H}_\mathrm{XY} = -\frac{1}{4L} \sum_{i,j} \left( \tilde{\sigma}^+_i \tilde{\sigma}^-_j + \tilde{\sigma}^-_i \tilde{\sigma}^+_j \right) + \Delta \sum_i \sigma^z_i,
    \end{equation}
    and $\tilde{H}_\mathrm{drive}$ is a non-conserving drive term given by
    \begin{equation}
        \tilde{H}_\mathrm{drive} = -\frac{\chi}{4L} \sum_{i,j} \left( \tilde{\sigma}^+_i \tilde{\sigma}^+_j + \tilde{\sigma}^-_i \tilde{\sigma}^-_j \right),
    \end{equation}
    with $\tilde{\sigma}^+_i = (\tilde{\sigma}^x_i + i\tilde{\sigma}^y_i)/2$ the projected spin-ladder operators. While for $\chi = 1$, we recover the original Hamiltonian $\tilde{H}$, we consider the limit $\chi \ll 1$ such that the drive term can be treated perturbatively. We will show below that the limit of $\chi=1$ can be qualitatively explained based on similar arguments.

    The naming of $\tilde{H}_\mathrm{drive}$ becomes clear upon moving to a rotating frame defined by the transformation with respect to the $\Delta \sum_i \sigma^z_i$ term. In this frame, the Hamiltonian becomes
    \begin{equation}
        \tilde{H}^r(\chi) = \tilde{H}^r_\mathrm{XY} + \tilde{H}^r_\mathrm{drive}(\chi),
    \end{equation}
    where
    \begin{equation}
        \tilde{H}^r_\mathrm{XY} = -\frac{1}{4L} \sum_{i,j} \left( \tilde{\sigma}^+_i \tilde{\sigma}^-_j + \tilde{\sigma}^-_i \tilde{\sigma}^+_j \right),
    \end{equation}
    \begin{equation}
        \tilde{H}^r_\mathrm{drive}(\chi) = -\frac{\chi}{4L} \sum_{i,j} \left( e^{+4i\Delta t} \tilde{\sigma}^+_i \tilde{\sigma}^+_j + e^{-4i\Delta t} \tilde{\sigma}^-_i \tilde{\sigma}^-_j \right).
    \end{equation}

    We begin by noting that the N\'eel state is an eigenstate of the unperturbed Hamiltonian, $\tilde{H}^r_\mathrm{XY} \ket{Z_2} = -0.5 \ket{Z_2}$.
    To analyze the spectrally nearby excitations, we apply the Holstein–Primakoff transformation in the dilute limit:
    \begin{equation}
        \tilde{\sigma}^-_i \approx 2 a^\dagger_i,
    \end{equation}
    while restricting the dynamics to the occupied sublattice due to the blockade. Upon transforming to momentum space, we obtain
    \begin{equation}\label{eq:H_xy_magnon}
        \tilde{H}^r_\mathrm{XY} \approx -\frac{1}{2} - a^\dagger_{\boldsymbol{k}=0} a_{\boldsymbol{k}=0} ,
    \end{equation}
    which shows that the single-magnon gap above the N\'eel state is $\epsilon_1 = -1$. The presence of only the zero-momentum mode in Eq.~\eqref{eq:H_xy_magnon} reflects the long-range character of the interactions. The negative magnon energy indicates that the N\'eel state is dynamically unstable, although this instability cannot be accessed through the spin-conserving part $\tilde{H}^r_\mathrm{XY}$ alone.
    
    We now examine the effect of the drive term. In bosonic language, and again restricting to the occupied sublattice, it becomes
    \begin{equation}
        \tilde{H}^r_\mathrm{drive}(\chi) \approx -\frac{\chi}{2} e^{-4i\Delta t} a^\dagger_{\boldsymbol{k}=0} a^\dagger_{\boldsymbol{k}=0} + \mathrm{h.c.}
    \end{equation}
    For small $\chi$, this term can be treated perturbatively. According to Fermi’s golden rule, $\tilde{H}^r_\mathrm{drive}$ generates magnon pairs at zero momentum if the following resonance condition is met:
    \begin{equation}\label{eq:D_resonance}
        -4\Delta_\mathrm{res} \approx 2\epsilon_1 \quad \rightarrow \quad \Delta_\mathrm{res} \approx -0.5.
    \end{equation}
    Given the connection between magnon population and entropy in long-range systems (cf.~Eq.~\eqref{eq:S_logn_logt}), this implies that entropy production should exhibit a peak at the resonance frequency. As shown in Fig.~\ref{fig:S_vs_t_deform}, the prediction of Eq.~\eqref{eq:D_resonance} agrees well with the numerically observed behavior of the deformed Hamiltonian with $\chi = 0.1$.
    
    Likewise, the enhancement of entanglement growth in the original model ($\chi=1$) can be attributed to the same resonance mechanism. However, in that case the resonance is broader due to higher-order processes and magnon interactions. Consequently, the increase of growth rate is more gradual compared to the sharply defined profile in case of $\chi \ll 1$.

    \section{Experimental realization of the model}\label{sec:experimental}
    In the following, we provide a brief overview on how the model defined in Eqs.~\eqref{eq:H0}-\eqref{eq:H_ac} could be realized in an upcoming experimental platform based on Rydberg arrays coupled to single-mode optical resonators.
    
Specifically, we consider a setting in which two stable hyperfine ground states in alkali atoms are coupled off-resonantly to Rydberg states via Rydberg dressing~\cite{Zeiher_Rydberg2016} and at the same time to an optical resonator with two pump fields to realize a Dicke model~\cite{Baden_DickeRaman2014}. This setting and the mapping of the quantum optical model to the many-body system studied in this work was analyzed in detail in Ref.~\cite{Gelhausen_Buchhold_Rosch_Strack_2016}.

For the Ising interactions implemented via Rydberg dressing, we assume that the atoms can be coupled with a Rabi frequency of $\Omega/2\pi = 10\,$MHz at a detuning of $\Delta/2\pi = 20\,$MHz to a Rydberg state with a decay time of $\tau = 50\,\mu$s. For these parameters, the effective lifetime of the system is $\tau_\mathrm{eff}=4\tau\Delta^2/\Omega^2 = 800\,\mu$s and the Rydberg-dressed interaction $V=\Omega^4/8\Delta^3= 2\pi \times 156\,$kHz. This leads to a quality factor of $V\tau_\mathrm{eff} = 125$. We note that these parameters can be flexibly adjusted by changing the effective admixture to the Rydberg state, e.g. by increasing $\Delta$.

The coupling to the resonator with two coherent pump fields naturally leads to the emergence of the Dicke model~\cite{Baden_DickeRaman2014}, as required by Eq.~\eqref{eq:H_ac}. Adiabatic elimination of the cavity field leads to effective interactions accompanied with collective dissipation via photons leaking from the cavity, as well as free-space scattering due to off-resonant Raman coupling. It can be straightforwardly shown that the collective cooperativity sets the ratio between cavity-induced interaction and scattering rate to free space in this setting.  {For near-concentric cavities, single-atom cooperativities of about $C=4g_0^2/\kappa\gamma \approx 30$ are realistic for cavity parameters $(g_0, \kappa, \gamma)\approx 2\pi \times (2, 0.1, 6)$MHz, where $g_0=g/\sqrt{L}$ is the single atom-cavity coupling. Assuming that about $L=50$ atoms can be placed within the cavity mode, we have $g\approx 2\pi \times 14$ MHz, and the collective cooperativity of $CL = 1500$ allows for a significant number of coherent processes before dissipation appears.}

    \section{Perspectives}

   On the theoretical side, a natural next step is to extend the present framework to higher dimensions, particularly to two-dimensional arrays of Rydberg atoms in cavities, which are already within experimental reach. From a computational standpoint, analyzing the resulting models will likely require scalable approaches such as tensor network methods, or variational wavefunction techniques inspired by those developed in Ref.~\cite{Ho_scars2019} for PXP scars. 
   
   Another important direction is to incorporate the full tail of the Rydberg interaction beyond nearest neighbors. In pure Rydberg systems, it has been shown that the inclusion of these effects can stabilize exotic phenomena such as emergent anyonic excitations~\cite{Lesanovsky_Fibonacci2012}. It is an intriguing open question whether similarly nontrivial excitations can arise in the cavity-coupled setting considered here. Furthermore, our analysis neglects dissipation, which is unavoidable in realistic experimental platforms. However, as discussed in Section~\ref{sec:experimental}, estimates suggest that dissipation plays a limited role on experimentally accessible timescales, allowing the phenomena described in this work to be observed over sufficiently long time windows before decoherence becomes significant. Nonetheless, establishing the robustness of the non-thermal features and dynamical signatures identified here in the presence of loss remains an important open question for future theoretical and experimental studies.  {Related to this latter point,  we emphasize that in any realistic implementation, scar dynamics will emerge within a prethermal time window, prior to the onset of corrections to the idealized infinite-Rydberg interaction limit, which ultimately lead to deviations from our predictions.   This indirectly indicates that the scar dynamics identified in our work is not a fine-tuned phenomenon, but rather a robust feature that can arise under a broad range of experimental conditions, governing the dynamics at least up to intermediate times, and potentially beyond depending on the operational capabilities of the optical cavity and Rydberg arrays coupled to them. }

 {From a broader perspective, our work opens a new research direction at the interface of many-body quantum optics and kinetically constrained dynamics. A first natural question is how the Dicke phase transition is reshaped by the presence of Rydberg blockade. While here we have focused on non-equilibrium entanglement dynamics, understanding criticality in this setting requires probing magnetization dynamics and correlation functions — quantities directly accessible in experiments. Given the foundational role of the Dicke transition in phenomena ranging from superradiance and collective light–matter synchronization to cavity-mediated quantum phases and precision metrology~\cite{Kirton_Dicke2019}, its reformulation in the presence of kinetic constraints may lead to a broad set of applications. For instance, one could envision forms of super-radiant phases of matter where certain atomic configurations can collectively radiate   preferentially  compared to others due to constraints, and break the robust mean-field paradigm of Dicke transitions. 
A second natural extension is to consider a constrained version of the Tavis–Cummings model, obtained by removing counter-rotating terms while retaining Rydberg blockade. In this case, the conservation of total excitation number introduces an additional mechanism for slow relaxation, potentially stabilizing quantum correlations against internal dephasing. Since the unconstrained Tavis–Cummings model underpins spin squeezing~\cite{ma2011quantum}, it is particularly compelling to investigate how blockade-induced constraints deform these states beyond the collective-spin manifold, possibly enabling new forms of metrologically useful, non-Gaussian entanglement.
Finally, kinetic constraints may profoundly modify dissipative processes. Upon eliminating the cavity field in Eq.~\eqref{eq:H_tilde_ac}, collective emission channels (typically described by jump operators such as $L=S_x$) become dressed by local projectors enforcing the blockade ($L=S_x \to \tilde{S}_x$). This raises fundamental questions about whether constrained dissipation is a form of more benign  decoherence compared to its unconstrained counterpart~\cite{gross1982superradiance}, alter relaxation pathways, or even provide a resource for engineering entangled dissipative dynamics. These three directions represent all equally interesting opportunities in the novel emerging field of kinetically constrained quantum optics.}

    On the experimental side, a key step is the implementation of the proposed setup, namely, coupling Rydberg transitions to a cavity mode while maintaining a sufficiently large lattice size to access the many-body regime. For one-dimensional systems aligned with the cavity axis, the elementary building blocks of tweezers arrays combined with optical cavities~\cite{Deist_cavityArray1_2022,Deist_cavityArray2_2022}, as well as the implementation of Rydberg-dressed interactions in lattices and tweezers~\cite{Zeiher_Rydberg2016,Zeiher_Rydberg2017,Steinert_Rydberg2023} have been shown before, and experimental setups combining both capabilities are underway. Extending this architecture to two-dimensional settings is feasible for near-concentric cavity designs with mode-sizes sufficiently large to accommodate multiple chains of coupled atoms within the optical modes~\cite{de2026realization}. Pure Rydberg systems in two dimensions are known to support a wide variety of phases, both conventional (e.g., density waves with different spatial patterns) and unconventional (e.g., quantum spin liquids), depending on the lattice geometry and interaction range. Embedding such platforms in optical cavities opens a direct path to exploring how cavity-mediated long-range interactions reshape this rich phase diagram. Furthermore, recent proposals suggest that cavity coupling may actively stabilize exotic quantum phases such as valence bond solids and spin liquids~\cite{JMQSL}. Realizing such states in a controlled, tunable cavity QED setting would represent a major advance in quantum simulation.

    \begin{acknowledgements}
    	We thank Hannes Bernien, Berislav Buca, Aashish Clerk, Nicol\'o Defenu,   Giuliano Giudici, Mark Oehlgrien, Hannes Pichler and Federica Surace, for useful discussions. This project has been supported by  QuantERA II Program that has received funding from the European Union’s Horizon 2020 research and innovation program under Grant Agreement No 101017733 ``QuSiED'' and by the DFG (project number 499037529), and in part by grant NSF PHY-2309135 to the Kavli Institute for Theoretical Physics (KITP). D.E.C. acknowledges support from the European Union, under European Research Council grant agreement No 101002107 (NEWSPIN); the Government of Spain (Severo Ochoa Grant CEX2019-000910-S [MCIN/AEI/10.13039/501100011033]); QuantERA II project QuSiED, co-funded by the European Union Horizon 2020 research and innovation programme (No 101017733) and the Government of Spain (European Union NextGenerationEU/PRTR PCI2022-132945 funded by MCIN/AEI/10.13039/501100011033); Generalitat de Catalunya (CERCA program and AGAUR Project No. 2021 SGR 01442); Fundació Cellex, and Fundació Mir-Puig. J.Z. acknowledges funding by Deutsche Forschungsgemeinschaft (DFG, German Research Foundation) under Germany’s Excellence Strategy-EXC-2111-390814868 and the Research Unit FOR5522, from the BMBF through the program “Quantum technologies - from basic research to market” (SNAQC, Grant no. 13N16265). R.M. was supported by the Deutsche Forschungsgemeinschaft under grants FOR 5522 (project-id 499180199) and the cluster of excellence ct.qmat (EXC 2147, project-id 390858490).
    \end{acknowledgements}

    \appendix

    \section{Derivation of the effective Hamiltonian}\label{app:H_eff}

We present a derivation of the effective Hamiltonian in Eq.~\eqref{eq:H_PXP2} starting from the microscopic model in Eq.~\eqref{eq:H0}. We begin in the limit in which light and matter are decoupled ($g=0$), and subsequently increase $g$ to a finite but small value. For $g=0$, the cavity spectrum consists of equally spaced levels separated by $\omega_c$, while the atomic spectrum features level separations of $2\Delta$, $2\Delta+V$, and $2\Delta+2V$. These correspond, respectively, to atomic excitations at sites with no excited neighbors, one excited neighbor, and two excited neighbors.

The light-matter coupling term $H_{\mathrm{ac}}$ connects sectors of the Hilbert space with different numbers of excitations. It induces the following processes, together with their corresponding excitation energies:
\begin{enumerate}
    \item Photon emission from an excited atom without excited neighbors,
    \begin{equation}
        E_f - E_i = \omega_c - 2\Delta ,
    \end{equation}
    \item Photon emission from an excited atom with one excited neighbor,
    \begin{equation}
        E_f - E_i = \omega_c - 2\Delta - V ,
    \end{equation}
    \item Photon emission from an excited atom with two excited neighbors,
    \begin{equation}
        E_f - E_i = \omega_c - 2\Delta - 2V ,
    \end{equation}
\end{enumerate}
as well as their corresponding absorption processes. Within perturbation theory, the importance of each process at finite $g$ is governed by the control parameter $g \equiv \tilde{g}/|E_f - E_i|$. For $\tilde{g}\ll 1$, the associated process contributes only weakly to the system dynamics and can be accurately captured within low-order perturbation theory. In the following, we analyze the resulting effective descriptions in different parameter regimes.

\subsection{Strong Rydberg regime}

In the extreme limit $V/\omega_c \to \infty$, the Hilbert space decomposes into subspaces characterized by identical configurations of simultaneously excited neighbors, while the positions of isolated excitations may vary within each subspace. In this regime, $H_{\mathrm{ac}}$ can only flip spins which have no excited neighbors. The light-matter coupling is therefore projected onto
\begin{equation}\label{eq:H_tild_ac_app}
    H_{\mathrm{ac}} \to \tilde{H}_{\mathrm{ac}}
    = -\frac{g}{\sqrt{L}} \sum_i (a + a^\dagger)\,\tilde{\sigma}^x_i ,
\end{equation}
as given in Eq.~\eqref{eq:H_tilde_ac}.

Taking the additional limit $\omega_c/\Delta \to \infty$ suppresses processes of type (1), leading to trivial dynamics. For large but finite $\omega_c/\Delta$, the cavity mode can be eliminated using second-order perturbation theory. This yields the effective interaction
\begin{equation}
    H_J
    = -\frac{J}{L}
      \sum_{i,j} \tilde{\sigma}^x_i \tilde{\sigma}^x_j ,
\end{equation}
where 
$J=g^2/2\omega_c$, in agreement with Eq.~\eqref{eq:H_xx}. Upon further decreasing $\omega_c$ such that $|\omega_c - 2\Delta| \lesssim g$, one can apply the rotating-wave-approximation to Eq.~\eqref{eq:H_tild_ac_app} to get
\begin{equation}
    \tilde{H}_\mathrm{ac} \approx -\frac{g}{2\sqrt{L}}\sum_i \big( a \tilde{\sigma}^+_i + a^\dagger \tilde{\sigma}^-_i \big).
\end{equation}
 Moreover, cavity losses become relevant in this regime, as they contribute to the atomic decay rate as
\begin{equation}
    \gamma \propto \frac{g^2 \kappa}{(\omega_c-2\Delta)^2+\kappa^2},
\end{equation}
which reaches its maximum when $\omega_c \approx 2\Delta$. A faithful description of this regime requires retaining both atomic and photonic degrees of freedom, which severely limits the feasibility of exact diagonalization.

\subsection{Moderately strong Rydberg regime}
When $\infty > V/\omega_c \gg 1$, processes of type (2) and (3) can influence the dynamics through virtual transitions, i.e., at higher orders in perturbation theory. In this case, second-order perturbation theory generates additional contributions to the effective Hamiltonian after adiabatically eliminating photons:
\begin{align}
    \tilde{H}_1 &=
    \frac{J'}{4L}
    \sum_{i,j} \Big[
    (n_{i-1} - n_{i+1})^2 \sigma^+_i\,
    (n_{j-1} - n_{j+1})^2 \sigma^-_j + (i\leftrightarrow j)\Big] ,
    \label{eq:H_til_1} \\
    \tilde{H}_2 &=
    \frac{J''}{4L}
    \sum_{i,j}\Big[
    n_{i-1} \sigma^+_i n_{i+1}\,
    n_{j-1} \sigma^-_j n_{j+1} + (i\leftrightarrow j)\Big] ,
    \label{eq:H_til_2}
\end{align}
where $J'=g^2/2(V-\omega_c)$ and $J''=g^2/2(2V-\omega_c)$. These terms allow the positions of simultaneously excited neighbors to change while conserving their total number. Consequently, the Hilbert space can be decomposed into sectors with fixed numbers of simultaneously excited neighbors. These sectors remain dynamically decoupled, and the evolution takes place entirely within the sector determined by the initial state.

This approach remains technically advantageous, as the cavity degrees of freedom are eliminated and the relevant Hilbert space is still significantly smaller than the full $2^L$-dimensional atomic Hilbert space. Finally, we note that the couplings in Eqs.~\eqref{eq:H_til_1} and~\eqref{eq:H_til_2} are antiferromagnetic. They compete with the ferromagnetic interaction in $H_J$, thereby introducing frustration that may give rise to novel phases. We emphasize that the appearance of these terms is a distinctive feature of our model and has no analogue in the standard PXP model. A detailed investigation of their consequences is beyond the scope of the present work and is left for future studies.
    
	\bibliography{Refs}

\end{document}